\documentclass[a4paper,fleqn]{cas-dc}

\usepackage[numbers,sort&compress]{natbib}
\usepackage{lipsum}
\def\tsc#1{\csdef{#1}{\textsc{\lowercase{#1}}\xspace}}
\tsc{WGM}
\tsc{QE}

\usepackage{url}

\usepackage[inline]{enumitem} %
\usepackage{multirow}
\usepackage{xcolor}
\usepackage{colortbl}
\usepackage{fancyvrb}
\usepackage{fvextra}
\usepackage{xspace}
\usepackage{flushend}
\usepackage{listings}
\usepackage{etoolbox}
\usepackage{tabularx}
\usepackage{pifont}

\makeatletter
\preto\lstlisting{\def\@captype{table}}
\makeatother

\usepackage{lipsum}

\usepackage{xspace}
\usepackage{makecell}
\newcommand{\secref}[1]{Section~\ref{#1}\xspace}
\newcommand{\figref}[1]{Figure~\ref{#1}\xspace}
\renewcommand{\tabref}[1]{Table~\ref{#1}\xspace} %

\newcommand{\ie}{\emph{i.e.},\xspace}
\newcommand{\eg}{\emph{e.g.},\xspace}

\usepackage{lipsum}
\usepackage{amsmath}

\usepackage{pgfplots} %
\usepackage[inkscapelatex=false]{svg} %
\usepackage{bbding} %
\usepackage[ruled,linesnumbered,vlined]{algorithm2e} %
\usepackage{amsthm} %
\newtheorem*{lemma}{Observation}

\usepackage[most]{tcolorbox}
\tcbset {
	base/.style={
		arc=0mm, 
		bottomtitle=0.5mm,
		boxrule=0mm,
		colbacktitle=black!10!white, 
		coltitle=black, 
		fonttitle=\bfseries, 
		left=2.5mm,
		leftrule=1mm,
		right=3.5mm,
		title={#1},
		toptitle=0.75mm, 
	}
}
\definecolor{brandblue}{rgb}{0.34, 0.7, 1}
\newtcolorbox{mainbox}[1]{
	colframe=brandblue, 
	base={#1}
}
\newtcolorbox{subbox}[1]{
	colframe=black!30!white,
	base={#1}
}

\usepackage{ifthen}
\newboolean{showcomments}
\setboolean{showcomments}{true}  %
\ifthenelse{\boolean{showcomments}}
{\newcommand{\nnote}[2]{
		\fbox{\bfseries\sffamily\scriptsize#1}
		{\sf\small$\blacktriangleright$\textit{#2}$\blacktriangleleft$}
	}
}
{\newcommand{\nnote}[2]{}
	
}

\usepackage{lipsum}
\usepackage{amsmath}

\newcommand\notsotiny{\@setfontsize\notsotiny\@vipt\@viipt}

\begin{document}
\let\WriteBookmarks\relax
\def\floatpagepagefraction{1}
\def\textpagefraction{.001}

\shorttitle{Converting \textsc{RegExes} into Dynatrace Pattern Language}    

\shortauthors{Fragner et al.}  

\title [mode = title]{Lost in Translation? Converting \textsc{RegExes} for Log Parsing into Dynatrace Pattern Language}

\author[1]{Julian Fragner}

\cormark[1]

\ead{julian.fragner@dynatrace.com}

\affiliation[1]{organization={Dynatrace Austria GmbH},
	city={Klagenfurt},
	country={Austria}}

\affiliation[2]{organization={University of Klagenfurt},
	city={Klagenfurt},
	country={Austria}}

\author[2]{Christian Macho}[orcid=0000-0001-8182-7277]
\ead{christian.macho@aau.at}
\ead[url]{https://mitschi.github.io/}

\author[1]{Bernhard Dieber}[orcid=0000-0002-0450-8990]
\ead{bernhard.dieber@dynatrace.com}
\ead[url]{https://www.bernharddieber.com/}

\author[2]{Martin Pinzger}[orcid=0000-0002-5536-3859]
\ead{martin.pinzger@aau.at}
\ead[url]{https://pinzger.github.io/}

\cortext[1]{Corresponding author}

\begin{abstract}
Log files provide valuable information for detecting and diagnosing problems in enterprise software applications and data centers. Several log analytics tools and platforms were developed to help filter and extract information from logs, typically using regular expressions (\textsc{RegExes}). Recent commercial log analytics platforms provide domain-specific languages specifically designed for log parsing, such as Grok or the Dynatrace Pattern Language (DPL). However, users who want to migrate to these platforms must manually convert their \textsc{RegExes} into the new pattern language, which is costly and error-prone. 

In this work, we present \textsc{Reptile}, which combines a rule-based approach for converting \textsc{RegExes} into DPL patterns with a best-effort approach for cases where a full conversion is impossible. Furthermore, it integrates GPT-4 to optimize the obtained DPL patterns. The evaluation with 946 \textsc{RegExes} collected from a large company shows that \textsc{Reptile} safely converted 73.7\% of them. The evaluation of \textsc{Reptile}'s pattern optimization with 23 real-world \textsc{RegExes} showed an F1-score and MCC above 0.91. These results are promising and have ample practical implications for companies that migrate to a modern log analytics platform, such as Dynatrace.
\end{abstract}

\begin{keywords}
Log parsing\sep Regular expressions\sep Dynatrace Pattern Language\sep Log pattern conversion\sep Log pattern optimization
\end{keywords}

\maketitle

\section{Introduction}
\label{sec:introduction}

Enterprise software applications and data centers become increasingly complex, which in turn increases the effort of operating them~\cite{Zhang2023}. 
Often, seamless execution or agreed performance of software systems must be guaranteed~\cite{He2017}. 
In large organizations, dedicated teams are assigned to monitor software execution and identify system events at runtime. %
Log files are seen as a primary source for problem diagnosis~\cite{Fu2014}. %
They provide valuable runtime information of applications, which can be used for maintenance, troubleshooting, anomaly and problem detection, failure prediction, root cause analysis, performance diagnosis, and security threat detection~\cite{Zhang2023, He2017, Debnath2018, Le2023}.

To avoid downtimes and meet service-level agreements (SLAs) of business applications and data centers, it is important to analyze log files quickly and accurately~\cite{He2017, Debnath2018}. 
However, with the increasing size and number of software systems in an organization, the volume and variety of logs grow~\cite{Zhang2023, He2017}.
This makes manual inspection of log files practically infeasible~\cite{Haibo2018, Vaarandi2014, Hamooni2016}. 
Therefore, commercial log analytics tools and platforms, such as \textit{Splunk}\footnote{\url{https://www.splunk.com/}}, \textit{ElasticSearch}\footnote{\url{https://www.elastic.co/elasticsearch}}, \textit{Datadog}\footnote{\url{https://www.datadoghq.com/}}, and \textit{Dynatrace}\footnote{\url{https://www.dynatrace.com/}},
emerged to help filter and extract information from logs~\cite{Zhang2023}.

As no standardized format and official logging guidelines exist~\cite{Zhang2023,Fu2014}, log messages are mostly unstructured and vary widely in content and format.
To extract information from such heterogeneous logs, most log analytics platforms require human involvement to define log formats using customized patterns~\cite{Debnath2018}.
Regular expressions (\textsc{RegExes}) are commonly used for this task~\cite{Zhang2023, He2017, Splunk2023b}. %
However, creating correct \textsc{RegExes} can be challenging, time-consuming, and error-prone, and requires specialized skills and experience~\cite{He2017, Zhong2018b, Bartoli2012, Bartoli2014, Bartoli2016, Friedl2006, Li2021, Ye2020, Chen2020, Li2020, Davis2019, Michael2019, Zhang2023b}. 
Additionally, the \textsc{RegEx} syntax is often difficult to comprehend~\cite{Friedl2006, Li2020, Michael2019, Chapman2017}, making pattern maintenance cumbersome and error-prone~\cite{Zhang2023, He2017}.

To address these drawbacks, some providers utilize domain-specific languages (DSL) specifically designed for log parsing. 
For example, \textit{ElasticSearch}~\cite{Elastic2024b} and \textit{Datadog}~\cite{Datadog2024b} use \textit{Grok}, which builds on top of \textsc{RegExes}, while \textit{Dynatrace} uses the proprietary \textit{Dynatrace Pattern Language}~(DPL)~\cite{Dynatrace2024b,Dynatrace2024c}. 
These languages are generally more user-friendly than \textsc{RegExes} and require less technical background from the user. 
For example, they provide predefined matchers for commonly occurring log data, such as timestamps and IP addresses, which enhances efficiency of pattern creation and reduces the risk of human error.

\figref{lst:dpl-regex} demonstrates these advantages by comparing a sample log pattern in \textsc{RegEx}, Grok, and DPL syntax respectively.
Line 2 shows the \textsc{RegEx} that extracts the IP address, matches any text followed by at least one space, and finally extracts the response code from a log entry.
Lines 5 and 8 show its Grok and DPL counterparts, respectively, which are easier to read than the \textsc{RegEx} as they are shorter and contain fewer special characters.
The DPL additionally separates its individual matchers by spaces.
\begin{figure}
\caption{\textsc{RegEx} example and corresponding Grok and DPL pattern examples.}
    \label{lst:dpl-regex}
    \begin{Verbatim}[numbers=left,frame=single, xleftmargin=4pt, numbersep=0.2em]
// RegEx
(?<addr>\d{1,3}\.\d{1,3}\.\d{1,3}\.\d{1,3}).*\s+(?<rc>\d{3})

// Grok counterpart

// DPL counterpart
IPv4:addr LD SPACE+ INT:rc
    \end{Verbatim}
    
\end{figure}
However, the use of different pattern languages creates problems when users want to migrate from one log analytics provider to another. 
The manually created patterns must then be manually translated into the target language which is costly and error-prone. %

In this work, we present \textsc{Reptile}\footnote{\underline{R}egular \underline{E}xpression \underline{P}attern \underline{T}ranslation \underline{I}nto \underline{L}anguage \underline{E}quivalents}, an approach that converts \textsc{RegExes} into DPL patterns. 
We choose \textsc{RegExes} as the source language, as it is widely used by different log analytics and monitoring solutions as mentioned above and DPL as the target language, because this work was done in collaboration with Dynatrace which is a leading company in providing software observability solutions~\cite{gartner2024}.
\textsc{Reptile} facilitates automatic generation of DPL patterns purely from \textsc{RegExes} whenever possible. It also provides best-effort conversions and only involves the user in cases for which automatic translation is not possible.
As a first step towards \textsc{Reptile}, we investigate the usage of \textsc{RegExes} in practice to understand how practitioners use them. The corresponding research question is: 

\begin{itemize}
    \item[{\textbf{RQ1}}\label{rq:1}] \textbf{Which \textsc{RegEx} features are frequently used in practice?}
\end{itemize}
We studied a real-world dataset from a large organization from the retail industry and extracted 946 different \textsc{RegExes}. 
For each \textsc{RegEx}, we tallied the used \textsc{RegEx} features. We found that \textit{Named capturing group}, \textit{Greedy quantifier}, and \textit{Literal character} were the top-3 most used features ($\hookrightarrow$~\secref{sec:feature-count-results}). These findings laid the foundation for developing \textsc{Reptile} that we evaluated in our second research question:

\begin{itemize}
    \item[{\textbf{RQ2}}\label{rq:2}] \textbf{How effective is \textsc{Reptile} in converting \textsc{RegExes} into DPL patterns?}
\end{itemize}

The evaluation shows that our approach increases the percentage of safely convertible \textsc{RegExes} from initially less than 2\% to more than 73\% ($\hookrightarrow$~\secref{sec:eval-conversion}) in a real-world dataset.
Moreover, 897,000 generated test cases show that our approach retains the original \textsc{RegEx} semantics. 
These results show that our approach supports users in safely converting the majority of \textsc{RegExes} into DPL. Next, we investigate the usage of \mbox{GPT-4} to further optimize the DPL patterns by introducing high-level matchers. This leads to our third research question:

\begin{itemize}
    \item[{\textbf{RQ3}}\label{rq:3}] \textbf{What is the prediction accuracy of \textsc{Reptile}'s pattern optimization?}
\end{itemize}

The results with GPT-4 and 23 real-world \textsc{RegExes} show high prediction performance achieving an average F1-score and MCC above 0.91 across five high-level DPL matchers ($\hookrightarrow$~\secref{sec:eval-optimization}). This means that the majority of high-level matchers proposed by GPT-4 are correct. %

In summary, our work makes the following contributions: %
\begin{itemize}
\setlength\itemsep{0em}
    \item A study of most frequently used \textsc{RegEx} features for log parsing in practice, analyzing more than 900 real-world \textsc{RegExes} ($\hookrightarrow$~\secref{sec:feature-count-results}).
    \item Novel strategies to identify non-backtracking \textsc{Reg\allowbreak Exes}, enabling conversion of greedy and lazy quantifiers, which are the most frequently used features unsupported in DPL ($\hookrightarrow$~\secref{sec:conversion}).
    \item A two-phase approach for automatic conversion of \textsc{RegExes} into DPL, comprising of a rule-based conversion ($\hookrightarrow$~\secref{sec:conversion}) and an optimization with GPT-4 ($\hookrightarrow$~\secref{sec:optimization}). %
\end{itemize}

Our results have ample practical implications for companies that migrate or migrated to a modern log analytics platform, such as Dynatrace. They can use \textsc{Reptile} to automatically and safely convert the majority (in our case 73\%) of their \textsc{RegExes} for log parsing to DPL patterns. Furthermore, they can use \textsc{Reptile} to optimize the converted DPL patterns making them easier to understand and maintain with high precision and recall.

The remainder of the paper is organized as follows.
First, \secref{sec:background} defines terms and concepts. Next, we study the most frequently used \textsc{RegEx} features in
\secref{sec:feature-count}.
\secref{sec:approach} presents the converter prototype \textsc{Reptile}. %
\secref{sec:evaluation} assesses the proportion of safely convertible \textsc{RegExes} and their correctness and evaluates the accuracy of high-level matcher prediction.
\secref{sec:relwork} situates the paper with respect to related work. %
\secref{sec:conclusion} discusses the main results, current limitations, and potential threats to the validity.
Finally, \secref{sec:conclusion} concludes the paper and envisions future work.

\section{Background}
\label{sec:background}

This section outlines the key features of \textsc{RegExes} ($\hookrightarrow$~\secref{sec:regex}) and DPL ($\hookrightarrow$~\secref{sec:dpl}). 
We only present the most relevant features for this study. 
For more detailed explanations, we refer the reader to standard works on \textsc{RegExes}~\cite{Friedl2006, Fitzgerald2012} and the public DPL documentation~\cite{Dynatrace2024b}.

\subsection{Regular Expressions (\textsc{RegExes})}
\label{sec:regex}

This section is based on the work of Friedl~\cite{Friedl2006} and provides details of the most relevant \textsc{RegEx} features.
Note that in this work, the term \textsc{RegExes} does not strictly refer to the theoretical concept of regular expressions (first introduced by Kleene~\cite{Kleene1956} in 1956) as a way to express a regular language~\cite{Chomsky1956, Linz2022}. 
Instead, it refers to the practical implementation of regular expressions that are commonly used in programming languages, standard libraries, and text-processing applications~\cite{Chen2020, Davis2019, Michael2019}.
These practical implementations often support \textit{non-regular} features that extend the capabilities of regular languages, such as backreferences and lookarounds~\cite{Li2020, Moseley2023}.
Our investigation is limited to the widespread \textsc{PCRE} (Perl-compatible regular expressions) flavor~\cite{Friedl2006, PCRE1997}. 
A comprehensive comparison and implementation of different flavors is beyond the scope of this work.

\subsubsection{Backtracking Quantifiers}
\label{sec:backtracking}

\begin{table}
\caption{Overview of greedy quantifiers available in \textsc{RegExes}}
\begin{tabular}{ l l l }
\toprule
Quantifier & Equivalent & Match \\
\midrule
\Verb|?| & \verb|{0,1}| & zero times or once \\
\Verb|*| & \verb|{0,}| & zero, one, or multiple times \\
\Verb|+| & \verb|{1,}| & once or multiple times \\
\verb|{x}| & \verb|{x,x}| & exactly x times \\
\verb|{x,y}| & - & min. x, max. y times \\
\verb|{x,}| & - & min. x times \\
\bottomrule
\end{tabular}
\label{tbl:greedy-quantifiers}
\end{table}

\begin{figure}
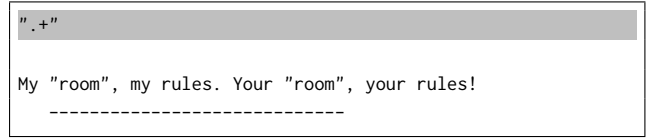

\caption{\textsc{RegEx} example: greedy quantifier}
\label{lst:greedy-quantifiers}
    \begin{Verbatim}[numbers=none,highlightlines=1,highlightcolor=lightgray,frame=single]
".+"

My "room", my rules. Your "room", your rules!
   -----------------------------
    \end{Verbatim}
    \vspace*{-4mm}
\end{figure}

Quantifiers are a central feature of \textsc{RegExes}.
They indicate that their preceding matcher can match optionally, or more than only once. 
See \tabref{tbl:greedy-quantifiers} for an overview of available quantifiers. 
Of all quantifiers, the most frequently used are the \textit{greedy} ones, which match as much of the input as possible. 
An example of this is shown in \figref{lst:greedy-quantifiers}, where the match result may seem counterintuitive.
The first line with gray background contains the \textsc{RegEx}. 
Below that is the input string to which the \textsc{RegEx} is applied, which we refer to as the \textit{target string}. 
The match result is indicated by dashes below the target string. 
The set of all strings accepted by a given \textsc{RegEx} is referred to as its \textit{defined language}.
We continue to use these typographical and terminological conventions hereinafter. %

Referring to the example in \figref{lst:greedy-quantifiers}, instead of matching the first \Verb|"room"| (as one might expect), the actual match is much longer. 
This is due to the use of the greedy quantifier~(\Verb|+|). 
First, the opening quotation mark (\Verb|"|) is matched followed by any character (\Verb|.|) at least once (\Verb|+|).
As the plus quantifier is greedy, it matches up to the end of the string. 
After that, the closing quotation mark in the \textsc{RegEx} must be matched. 
As a consequence, the \textsc{RegEx} engine \textit{backtracks} to a previous state where the last quotation mark in the target string was not matched yet.
This process is referred to as \textit{releasing} characters.
In the example, the characters already matched are released again one after the other (\ie first '\Verb|!|', then '\Verb|s|', '\Verb|e|', '\Verb|l|', etc.) until a quotation mark is found and an overall match is reported.

\begin{figure}
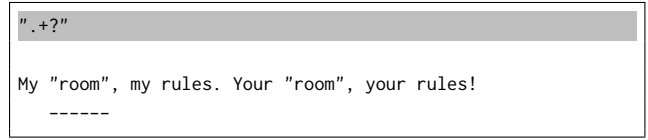

\caption{\textsc{RegEx} example: lazy quantifier}
\label{lst:lazy-quantifiers}
    \begin{Verbatim}[numbers=none,highlightlines=1,highlightcolor=lightgray,frame=single]
".+?"

My "room", my rules. Your "room", your rules!
   ------
    \end{Verbatim}
    \vspace*{-4mm}
\end{figure}

As demonstrated in the example above, greedy quantifiers may release characters that were already matched if necessary for an overall match to succeed. 
However, characters that are \textit{essential} to the quantifier are never released.
For example, the plus quantifier (\Verb|+|) in \figref{lst:greedy-quantifiers} would release all matched characters again if necessary, except the leftmost '\verb|r|', because the plus quantifier requires at least one match. 

All quantifiers presented in \tabref{tbl:greedy-quantifiers} can be turned into \textit{lazy} ones by appending a question mark. 
Their lazy equivalents are therefore \Verb|??|, \Verb|*?|, \Verb|+?|, and \Verb|{x,y}?| and they 
behave in the opposite way to greedy quantifiers, always matching as little as possible. 
In other words, they only match what is necessary and then hand over control to the next matcher. 
Only if the next matcher fails, the lazy matcher attempts another match.

Referring to the example in \figref{lst:lazy-quantifiers}, after the first quotation mark (\Verb|"|) matches, the lazy dot-matcher matches one character ('\Verb|r|') and then immediately hands over the control to the second '\Verb|"|'-matcher.
As the '\Verb|"|' and the next character in the target string '\Verb|o|' do not match, the control is handed back to the dot-matcher, which matches the '\Verb|o|'.
This procedure repeats until the second quotation mark in the target string is reached and an overall match can be reported.

\subsubsection{Non-Backtracking Quantifiers}
\label{sec:nonbacktracking}

Similar to the previous section, the greedy quantifiers presented in \tabref{tbl:greedy-quantifiers} can be turned into \textit{possessive} ones by appending a plus symbol (\Verb|+|). 
The possessive equivalents are \Verb|?+|, \Verb|*+|, \Verb|++|, and \Verb|{x,y}+|.
In contrast to matchers with greedy and lazy quantifiers, matchers with possessive quantifiers never backtrack (\ie never release characters once matched). 
\figref{lst:possessive-quantifiers} shows an example where the match result is the same as if a standard greedy quantifier was used.
However, for target strings where there is no match (for example, if the target string does not contain a colon), the \textsc{RegEx} engine does not need to backtrack and release already matched characters.
That is, a match failure can be reported earlier.
This can save memory and increase runtime performance, especially when such a \textsc{RegEx} is applied to many non-matching target strings.
This optimization can be applied because, in this example, a colon (\Verb|:|) will never appear among the already matched characters due to the \Verb|[a-z]| character class, making backtracking obsolete.

\begin{figure}
\caption{\textsc{RegEx} example: possessive quantifier}
    \label{lst:possessive-quantifiers}
    \begin{Verbatim}[numbers=none,highlightlines=1,highlightcolor=lightgray,frame=single]
^[a-z]++:

rules: 1) ... 2) ...
------
    \end{Verbatim}
    \vspace*{-4mm}
    \end{figure}
    
\begin{figure}
\caption{\textsc{RegEx} example: possessive quantifier pitfall}
    \label{lst:possessive-quantifiers-pitfall}
    \begin{Verbatim}[numbers=none,highlightlines=1,highlightcolor=lightgray,frame=single]
\d*+[0-9]

room number 345
(no match)
    \end{Verbatim}
    \vspace*{-4mm}
    
\end{figure}

Despite the potential benefits, it is recommended to use possessive quantifiers with caution~\cite{Friedl2006}. 
For example, in \figref{lst:possessive-quantifiers-pitfall}, the use of a possessive quantifier leads to a match failure. 
A greedy quantifier would match all three digits '\Verb|345|' one after the other and then release the '\Verb|5|' again so that the character class \Verb|[0-9]| can match. 
However, with a possessive quantifier, this last step does not happen leading to an overall match failure.

\subsection{Dynatrace Pattern Language}
\label{sec:dpl}

This section provides details on the most relevant DPL features.
It is based on the public DPL documentation~\cite{Dynatrace2024b} and discussions with the original DPL authors.
Note that DPL patterns are by default applied exactly once to the target string, \ie the DPL engine stops execution once it found a match for a given pattern.
In addition, matches are only attempted from the beginning of the target string. 
In other words, every DPL pattern is preceded by an implicit "beginning of line" matcher.

\subsubsection{Pattern Syntax}
\label{sec:dpl-syntax}

To match literal text in a target string, the required characters must be enclosed in double or single quotes within the DPL pattern.
If metacharacters (such as \Verb|*|, \Verb|[|, and \Verb|?|) are to be matched literally, they do not need to be escaped separately, except for the double and single quote and the backslash itself, which must be marked with a preceding backslash (\Verb|\"|, \Verb|\'|, and \Verb|\\|).
Note that literal matchers consisting of multiple letters are considered a single matcher in the DPL. 
For example, while \Verb|abc| represents three consecutive individual matchers in a \textsc{RegEx}, \Verb|"abc"| is treated as a single matcher in the DPL. 

The DPL provides so-called \textit{export names}, to assign names to parts of the pattern.
They are the counterpart to the named capturing groups in \textsc{RegExes}.
To export a matcher, a colon (\Verb|:|) is appended, followed by the actual export name.
It must start with a letter followed by alphanumeric characters~(letters, digits, and underscore). 
If a period (\Verb|.|) appears in the name, the entire export name must be enclosed in quotation marks. 
Export names can be applied to entire groups, or even individual matchers, and therefore do not necessarily require parentheses.
\tabref{tbl:feature-count} shows examples of the most common \textsc{RegEx} features and their equivalents in DPL syntax.

\subsubsection{Quantifier Semantics}

All the quantifiers listed in \tabref{tbl:greedy-quantifiers} are also supported by the DPL. 
Only the question mark (\Verb|?|) has slightly different semantics. It makes the entire matcher with its default quantifier optional, instead of forcing the matcher to match zero times, or exactly once. 
For example, \Verb|DIGIT?| makes the \Verb|DIGIT| matcher with its default quantifier \Verb|{1,4096}| optional, effectively translating to \Verb|DIGIT{0,4096}| (equivalent to \verb|\d{0,4096}| in \textsc{RegExes}). 
To exactly match zero or one digit~(like \verb|\d?|), the correct DPL pattern is \Verb|DIGIT{0,1}|.
Similarly, to exactly match one digit (like \verb|\d| in \textsc{RegExes}), the correct DPL pattern is \Verb|DIGIT{1}|.
In addition, the DPL supports the range quantifier \Verb|{,x}|, which is syntactic sugar for \Verb|{0,x}|, \ie matching minimum 0 and maximum \Verb|x| times.

The most critical difference to \textsc{RegExes}, however, is that despite having the same syntax as greedy quantifiers~($\hookrightarrow$~\secref{sec:backtracking}), their behavior is not greedy, but \emph{possessive}~($\hookrightarrow$~\secref{sec:nonbacktracking}). 
The only exception to this is when the quantifiers are applied to the \Verb|LD| or \Verb|DATA| matchers, which always exhibit \textit{lazy-like} behavior.
That is, they match any character up until their succeeding matcher, but (unlike a lazy quantifier in \textsc{RegExes}), they never expand their initial match.

To emphasize this, the DPL engine \textit{never backtracks} to a previous state, except when evaluating alternatives. 
Discussions with the original DPL authors revealed that this limitation exists for performance reasons, as backtracking can be an expensive operation.
In particular, for certain \textsc{RegExes}, so-called \textit{catastrophic backtracking} can occur, where the runtime complexity becomes super-linear, also known as \textit{runaway regular expressions}~\cite{Friedl2006, Fujinami2024, Li2020, Davis2019, Berglund_2014_catastrophicbacktracking}.

\subsubsection{High-Level Matchers}
\label{sec:high-level-matchers}

Additionally, the DPL provides a broad range of built-in matchers~\cite{Dynatrace2024m} for frequently occurring information in logs that do not have a direct equivalent in \textsc{RegExes}. 
We categorize them into two distinct groups:
\begin{enumerate*}[label=(\roman*)]
    \item matchers with a generic scope, \eg \verb|INT|, \verb|LONG|, \verb|DOUBLE|, and 
    \item matchers with a specific scope, \eg \verb|IPADDR|, \verb|CREDITCARD|, \verb|TIMESTAMP|
\end{enumerate*}.

Utilizing these matchers offers a number of advantages. 
Firstly, they improve the pattern by providing a layer of abstraction, rendering it more readable and maintainable.
This not only increases efficiency when writing patterns, but also reduces the risk of errors by eliminating the need to manually create potentially complex patterns. 
Secondly, they match specific values (\eg only matching valid credit card numbers), resulting in a more accurate pattern. 
Thirdly, a data type is associated with the matchers. 
For example, if a match of type \verb|INT| is exported, the exported value can be processed as an integer (\eg to easily check whether an HTTP response code is in the range of 200 to 300).
In contrast, \textsc{RegExes} export all matches as strings. 

\section{Preliminary Study}
\label{sec:feature-count}

In this section, we investigate which \textsc{RegEx} features are frequently used in practice, thereby establishing the foundation for our approach.
\secref{sec:dataset} presents the real-world dataset with 946 individual \textsc{RegExes} for log parsing.
The features used in these patterns are then counted and summarized in \secref{sec:feature-count-results}.

\subsection{Real-World Dataset}
\label{sec:dataset}

To identify which \textsc{RegEx} features are frequently used in practice (\hyperref[rq:1]{RQ1}), we study a real-world dataset. 
The dataset was provided by Dynatrace and consists of 5 CSV files with a total size of 3.2 MB.
It originates from a large organization in the retail industry which has migrated from Splunk to Dynatrace.\footnote{Please note, for reasons of confidentiality, neither the original dataset nor the extracted \textsc{RegExes} can be made publicly available. 
The examples presented in this work were selected to not contain sensitive information to ensure data privacy compliance.} 
The files contain various \textit{Search Processing Language}~(SPL) queries\footnote{Search Processing Language (SPL) queries are used to search, filter, and manipulate data within the Splunk platform~\cite{Splunk2023c}.}, including \textsc{RegExes} to parse query results. 

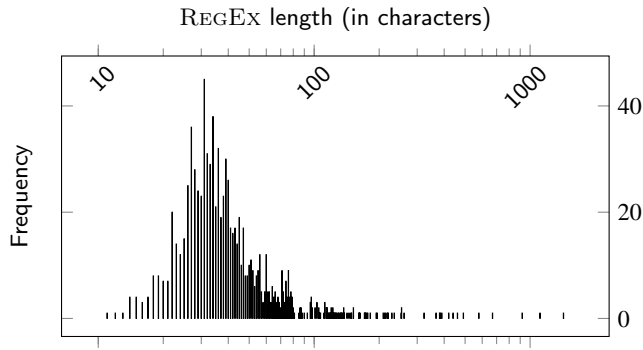
\begin{figure}
\caption{\textsc{RegEx} length distribution in the real-world dataset}
\label{fig:length-dist}
\centering
\begin{tikzpicture}
\begin{axis}[
    ybar,
    xmode=log, %
    xtick={10,100,1000}, %
    xticklabels={10,100,1000},
    xticklabel pos=upper,
    x tick label style={rotate=45,anchor=east,yshift=-10pt},
    title={\textsc{RegEx} length (in characters)},
    ylabel={Frequency},
    ylabel style={yshift=-20pt},
    yticklabel pos=right,
    width=250,
    height=150
]
\addplot [fill=gray, bar width=0.1pt] table [x=length, y=frequency, col sep=comma] {lst/pattern-lenghts.txt};
\end{axis}
\end{tikzpicture}
\end{figure}

For every query, we searched for the \Verb|rex| command~\cite{Splunk2023d} and extracted the contained \textsc{RegEx}. 
Because the \textsc{RegExes} were embedded in both SPL query and CSV format, the \textsc{RegExes} contained corresponding escapings, which were removed.
We obtained 2,701 \textsc{RegExes} from the original files. %
After eliminating duplicates, 946 distinct \textsc{RegExes} finally remained.
The \textsc{RegExes} range from 11 to 1,428 characters in length, with an average length of 56.5 characters.
\figref{fig:length-dist} shows the length distribution across all \textsc{RegExes}.
Note that the x-axis has a logarithmic scale.

For every \textsc{RegEx} in the dataset, we counted the features listed in \tabref{tbl:feature-count}.
To this end, a program written in TypeScript (TS) was implemented using the \textit{regexpp} parser library~\cite{regexpp2021} to analyze the \textsc{RegEx}.
First, match mode modifiers were removed so that the \textsc{RegEx} can be parsed by the \textit{regexpp} library.
Further, the following optimizations were applied:
\begin{itemize}
    \item Simplify character classes that contain a single element. Replace \Verb|[x]| by \Verb|x| if \Verb|x| is a single literal matcher or class shorthand.
    \item Simplify negated character classes that contain a single class shorthand. Replace \verb|[^\d]| by \Verb|\D| and vice-versa (\verb|[^\D]| by \verb|\d|) for all three kinds of class shorthands (\verb|\d|, \verb|\w|, and \verb|\s|).
\end{itemize}

Next, each \textsc{RegEx} was parsed, and the resulting abstract syntax tree (AST) was then traversed recursively.
For every intermediate or leaf node, the respective \textsc{RegEx} feature was counted.
For alternatives, we counted the number of pipe characters (\eg \Verb$(a|b)$ counts as 1, \Verb$(a|bc|d)$ counts as 2 and so forth).

\subsection{Results}
\label{sec:feature-count-results}

After analyzing all \textsc{RegExes}, the total feature count was obtained by summing up the counts within each \textsc{RegEx}.
\tabref{tbl:feature-count} presents the feature counts over all \textsc{RegExes}.
Column \textit{Total~\#} shows the total number of occurrences per feature in the entire dataset, while \textit{Affected~\#} counts the number of \textsc{RegExes} where the respective feature occurs at least once.
The table is sorted by the last column \textit{Affected~\%}, which shows the percentage of affected \textsc{RegExes}, calculated by dividing the number of affected \textsc{RegExes} by the total number of \textsc{RegExes} (946 entries).
For every feature, a \textsc{RegEx} example is given along with its corresponding DPL counterpart. 
The ten most common \textsc{RegEx} features are separated from the remaining ones by a horizontal line.
\textsc{RegEx} features that do not appear in \tabref{tbl:feature-count} (such as match boundaries, backreferences, and comments) did not appear in the entire dataset.
One exception are mode modifiers which had to be excluded as mentioned above.

\begin{table*}
\caption{Frequency of \textsc{RegEx} features (derived from Friedl~\cite{Friedl2006}) in the real-world dataset; \textsc{RegEx} features unsupported in DPL are marked (\ding{55}); \textsc{RegEx} features where conversion is possible with workarounds are marked (\ding{118}); remaining \textsc{RegEx} features can be directly converted to DPL.}
\label{tbl:feature-count}
\begin{tabular}{ l l l r r r }
\toprule
Feature Name & \textsc{RegEx} & DPL & Total \# & Affected \# & Affected \% \\
\midrule
Named capturing group & \Verb|(?<name>abc)| & \Verb|"abc":name| & 1517 & 945 & 99.9\% \\
Greedy quantifier (\ding{55}) & \Verb|a*| & \ding{55} & 2640 & 893 & 94.4\% \\
Literal character & \Verb|abc| & \Verb|"abc"| & 15983 & 891 & 94.2\% \\
Digit matcher & \Verb|\d| & \Verb|DIGIT{1}| & 825 & 384 & 40.6\% \\
Character representation & \Verb|\n| & \Verb|LF| & 1142 & 370 & 39.1\% \\
Dot-matcher & \Verb|.| & \Verb|LD| & 673 & 328 & 34.7\% \\
Character class (\ding{118}) & \Verb|[abc]| & \Verb|[abc]| & 402 & 233 & 24.6\% \\
Space matcher & \Verb|\s| & \Verb|SPACE{1}| & 627 & 185 & 19.6\% \\
Word matcher & \Verb|\w| & \Verb|WORD{1}| & 503 & 155 & 16.4\% \\
Negated character class (\ding{118}) & \Verb|[^abc]| & \Verb|[^abc]| & 282 & 143 & 15.1\% \\
\midrule
Lazy quantifier (\ding{55}) & \Verb|a*?| & \ding{55} & 177 & 63 & 6.7\% \\
Capturing group & \Verb|(ab)c| & \Verb|("ab")"c"| & 108 & 53 & 5.6\% \\
Line start & \Verb|^| & \Verb|BOS| & 45 & 45 & 4.8\% \\
Alternative & \Verb$a|bc$ & \Verb$("a"|"bc")$ & 85 & 41 & 4.3\% \\
Quantified group & \verb|(\s\w)*| & \Verb|ARRAY{SPACE{1} WORD{1}}*| & 72 & 40 & 4.2\% \\
Non-space matcher & \Verb|\S| & \Verb|NSPACE{1}| & 63 & 33 & 3.5\% \\
Non-capturing group & \Verb|(?:abc)| & \Verb|("abc")| & 44 & 30 & 3.2\% \\
Non-word matcher & \Verb|\W| & \Verb|[^a-zA-Z0-9]| & 13 & 10 & 1.1\% \\
Line end & \verb|$| & \verb|EOS| & 8 & 8 & 0.8\% \\
Optional group & \Verb|(abc)?| & \Verb|("abc")?| & 42 & 7 & 0.7\% \\
Non-digit matcher & \Verb|\D| & \Verb|[^0-9]| & 9 & 6 & 0.6\% \\
Positive lookahead & \Verb|(?=abc)| & \Verb|>>"abc"| & 3 & 3 & 0.3\% \\
Quantified named capturing group (\ding{55}) & \Verb|(?<name>abc)*| & \ding{55} & 4 & 2 & 0.2\% \\
Non-word boundary (\ding{118}) & \Verb|\B| & \ding{118} & 1 & 1 & 0.1\% \\
\bottomrule
\end{tabular}
\end{table*}

The prevalence of named capturing groups can be explained by their use in SPL for data extraction~\cite{Splunk2023b}.
The three \textsc{RegEx} features marked as~\ding{55} in \tabref{tbl:feature-count}, namely greedy quantifier, lazy quantifier, and quantified named capturing group, hinder a direct automatic translation. They are not supported by DPL.
Due to the low number of occurrences, we ignore quantified named capturing groups. For the other two, we observed that at least one of these quantifiers occurs in 930 \textsc{RegExes}. 
Three more \textsc{RegEx} features marked as~\ding{118}, namely character class, negated character class, and non-word boundary, are not directly supported by DPL, but can be emulated.\footnote{Character classes are compatible to DPL, with the exception of class shorthands within character classes. For example, the \textsc{RegEx} \Verb|[\d\w]| can be converted to \Verb|[0-9a-zA-Z_]| or \Verb$(DIGIT{1}|WORD{1})$ in DPL. Non-word boundaries (\Verb|\B|) can be emulated with lookarounds, which are supported in DPL.}

\begin{subbox}{Answer to \hyperref[rq:1]{RQ1}}
The top-10 \textsc{RegEx} features include named capturing groups, greedy quantifiers, (escaped) literals, (negated) character classes, non-negated class shorthands, and the dot-matcher.
\end{subbox}

The results from this study mean that only about 1.7\% of all \textsc{RegExes} from the original dataset can be converted directly into DPL patterns. %
The following section presents our \textsc{Reptile} approach that aims to increase this proportion.

\section{\textsc{Reptile} Approach}
\label{sec:approach}

\begin{figure}
\caption{Overview of the \textsc{Reptile} approach}
\label{fig:app-overview}
\centering
\includesvg[width=\columnwidth]{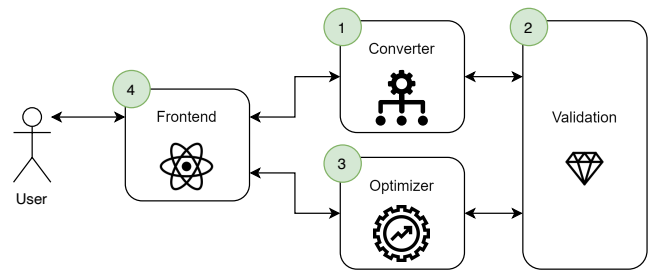}

\end{figure}

This section presents our \textsc{RegEx} converter approach \textsc{Reptile} (\underline{R}egular \underline{E}xpression \underline{P}attern \underline{T}ranslation \underline{I}nto \underline{L}ang\-uage \underline{E}quivalents).
\figref{fig:app-overview} shows an overview of \textsc{Reptile}.

In the first phase, the user submits a \textsc{RegEx} through the Frontend~(4), which is then forwarded to the Converter component~(1), where the basic rule-based conversion from \textsc{RegEx} to DPL happens.
Next, the converted DPL pattern is checked for correctness using the Validation~(2) component.
The converted pattern is presented in the Frontend~(4).
At this point, the user can already decide to accept the DPL pattern as final result. %

In an optional second phase, optimization of the DPL pattern can be initiated.
In this case, the basic DPL pattern is forwarded to the Optimizer~(3) component, which detects potential high-level DPL matchers. 
In the Frontend~(4), the user must then manually decide for each matcher whether the suggestion should be applied.
Optionally, the optimized pattern can be validated again.
The following sections provide detailed information on each individual component. 

\subsection{Rule-Based Conversion (Step 1)}
\label{sec:conversion}

The Converter component utilizes the \textit{regexpp} parser library~\cite{regexpp2021} to parse the input \textsc{RegEx} (similar as in \secref{sec:feature-count}).
The resulting AST is then traversed depth-first.
Every conversion step handling one AST node type incrementally extends the DPL pattern result.
For most \textsc{RegEx} features, a one-to-one mapping to DPL can be implemented, as shown in \tabref{tbl:feature-count}.
However, greedy and lazy quantifiers require special consideration.
To effectively convert greedy and lazy quantifiers from a \textsc{RegEx} to possessive quantifiers in DPL, it is necessary to identify situations where no backtracking is required.
In the following, we present these situations and our conversion strategies.

\subsubsection{Greedy Quantifiers}
We identify three cases where greedy quantifiers do not require backtracking and therefore can be converted to DPL.

\paragraph{Fixed Greedy Quantifier (FGQ).}
The first case of quantifiers without backtracking are quantifiers with fixed repetition, such as \Verb|{x,x}| or \Verb|{x}| in short. 
As mentioned in \secref{sec:backtracking}, greedy quantifiers may only release non-essential characters. 
For example, in the case of the quantifier \verb|{2,4}|, only the first two matched characters are essential, and the following two may be released again if needed. 
Similarly, for the plus quantifier (\verb|+|), the first matched character is essential, all following ones may be released again.
In the quantifier with fixed repetition \Verb|{x}| however, \textit{all x characters are essential}, and therefore, there is no backtracking, \ie no characters are released. 

\paragraph{Last Greedy Quantifier (LGQ).}
The second case, where no backtracking can happen, is when the greedily quantified matcher appears last in the \textsc{RegEx} (or last in any top-level alternative). %
For example, in \figref{lst:safe-greedy-last}, once the '\Verb|method=|' part matched, the following \Verb|[A-Z]*| part greedily matches uppercase letters until either
\begin{enumerate*}[label=(\roman*)]
    \item the end of the target string is reached, or
    \item a character is reached which does not match the expression, \ie is not an uppercase letter. 
\end{enumerate*}
This example shows the latter case.
However, in either case, the \Verb|[A-Z]*| never has to release any already matched characters again because no matcher comes afterwards that requires one of the matched characters.

\begin{figure}
\caption{Safe greedy quantifier: last matcher in a \textsc{RegEx}}
    \label{lst:safe-greedy-last}
    \begin{Verbatim}[numbers=none,highlightlines=1,highlightcolor=lightgray,frame=single]
method=[A-Z]*

method=POST, endpoint=https://...  
-----------
    \end{Verbatim}
    \vspace*{-4mm}
\end{figure}
\begin{figure}
\caption{Safe greedy quantifier: followed by non-intersecting literal matcher}
    \label{lst:safe-greedy-lit}
    \begin{Verbatim}[numbers=none,highlightlines=1,highlightcolor=lightgray,frame=single]
\d{1,3}x

789
(no match)
    \end{Verbatim}
\end{figure}
\begin{figure}
\caption{Unsafe greedy quantifier: followed by intersecting character class}
    \label{lst:unsafe-greedy-intersect}
    \begin{Verbatim}[numbers=none,highlightlines=1,highlightcolor=lightgray,frame=single]
\w+[a-z]

Hello-Muehlviertel!
-----
    \end{Verbatim}
\end{figure}
\begin{figure}
\caption{Unsafe greedy quantifier: followed by optional matcher and character class}
    \label{lst:unsafe-greedy-intersect-2}
    \begin{Verbatim}[numbers=none,highlightlines=1,highlightcolor=lightgray,frame=single]
\w+\s?[a-z]

Hello-Muehlviertel!
-----
    \end{Verbatim}
    
\end{figure}

\paragraph{Non-Intersecting Greedy Quantifier (NGQ).}
The third case where a matcher with a greedy quantifier can be safely converted is if its succeeding matcher matches different characters.
That is, the languages defined by a matcher and its successor do not intersect.
In set mathematics, detecting this condition is called the \textit{intersection \mbox{(non-)emptiness} problem}.
In the implementation of \textsc{Reptile}, we follow an existing approach~\cite{Su2023} and use the \textit{greenery} library~\cite{greenery2024} to check whether two matchers intersect.
First, consider the simplest case: a quantified matcher succeeded by a literal matcher as shown in \figref{lst:safe-greedy-lit}.
Here, the \verb|x| attempts to match, after the greedy \verb|d\{1,3}| reaches the end of the target string.
As the \verb|x| requires exactly one match, the digit matcher releases the last matched character \verb|9|.
The \verb|x| attempts to match the \verb|9| and fails, resulting in an overall failure.
It is important to realize that this backtracking step is always unnecessary, no matter to which target string the example \textsc{RegEx} is applied to.
The reason is that the two languages defined by \verb|\d{1,3}| and \verb|x| do not intersect.
While \verb|\d{1,3}| matches any numbers between \verb|0| and \verb|999| (including numbers with leading zeros, such as \verb|001|), the literal matcher \verb|x| solely matches the lowercase character \verb|x|.
Hence, no matter which digit is released by the digit matcher, it can never be consumed by the literal matcher \verb|x|.
In other words, the result does not change, when the greedy quantifier is replaced by a possessive one, making it safe to convert to DPL.
This insight is summarized in the following observation: %

\begin{lemma}\label{lemma:intersection}
Given a \textsc{RegEx} $P$, let $P = uv$, where $u$ and $v$ are subpatterns of $P$. Let $u$ be greedily quantified, and $v$ be non-optional. Let $L(u)$ and $L(v)$ be the languages defined by $u$ and $v$, respectively. If $L(u) \cap L(v) = \varnothing$, then the quantifier of $u$ can be replaced by a possessive quantifier without affecting $L(P)$.
\end{lemma}

\figref{lst:unsafe-greedy-intersect} shows an example where conversion is unsafe.
After the \verb|\w+| has matched \verb|Hello|, it passes the control to the final \verb|[a-z]|.
As this last matcher requires exactly one character, it forces the initial \verb|\w+| to release one character, finally matching the \verb|o|.
In summary, as the greedy \verb|\w+| and the succeeding matcher \verb|[a-z]| intersect, backtracking may be necessary. 
Hence, the \textsc{RegEx} cannot be safely converted to DPL.

The following special cases apply.
In case the successor is a group, the first element must be chosen from the group and checked for intersection.
If the group contains alternatives, intersection must be checked for the first element of every alternative.
In other words, the first element of every alternative must not intersect with the current matcher.
Note that \textit{successor} and \textit{first element} refer to the next non-optional matcher after the current matcher.
The reason is that optional matchers never cause the predecessor matcher to release characters that have already been matched, as they do not require any characters themselves.
Consider the example in \figref{lst:unsafe-greedy-intersect-2}.
Although \verb|\w+| and \verb|\s?| do not intersect, the result does not differ from the unsafe example in \figref{lst:unsafe-greedy-intersect}. 
Because \verb|\s?| is optional, and the target string does not contain a whitespace character, this matcher is skipped. 
The last matcher in turn initiates backtracking, thus forcing the initial \verb|\w+| to release one character.
In summary, because the greedy \verb|\w+| and the next non-optional matcher \verb|[a-z]| intersect, backtracking may be necessary and the \textsc{RegEx} cannot be safely converted to DPL.

\subsubsection{Lazy Quantifiers}

Most strategies for converging greedy quantifiers presented in the previous section also apply to lazy quantifiers.

\paragraph{Fixed Lazy Quantifier (FLQ).}
As explained in \secref{sec:backtracking}, lazy quantifiers expand their match if necessary for an overall match to succeed.
However, if a quantifier has fixed repetition (\Verb|{x,x}| or \Verb|{x}|) the quantifier can never expand its match beyond the initial match of length \verb|x|.
Hence, a matcher with such a quantifier can be safely converted to its possessive DPL counterpart.

\begin{figure}
\caption{Safe lazy quantifier: followed by non-intersecting literal matcher}
    \label{lst:safe-lazy-lit}
    \begin{Verbatim}[highlightlines=1,highlightcolor=lightgray,frame=single]
\d+?x$

78xx

    \end{Verbatim}
    
\end{figure}
\begin{figure}
\caption{Unsafe lazy quantifier: followed by intersecting character class}
    \label{lst:unsafe-lazy-intersect}
    \begin{Verbatim}[numbers=none,highlightlines=1,highlightcolor=lightgray,frame=single]
\w+?[a-z]

Hello-Lavanttal!
--
    \end{Verbatim}
    
\end{figure}
\begin{figure}
\caption{Safe lazy quantifier: dot-matcher followed by exactly one matcher}
    \label{lst:safe-lazy-dot}
    \begin{Verbatim}[numbers=none,highlightlines=1,highlightcolor=lightgray,frame=single]
.+?!

Hello! Zillertal!
------
    \end{Verbatim}
    
\end{figure}

\paragraph{Non-Intersecting Lazy Quantifier (NLQ).}
Next, a matcher with a lazy quantifier can be safely converted if it does not intersect with its next non-optional succeeding matcher as shown in Figure~\ref{lst:safe-lazy-lit}.
Here, \verb|\d+?| first matches the minimum number of required characters, namely the character \verb|7|.
Control is passed on to the \verb|x| matcher, which cannot match the next character in the target string (\verb|8|), forcing \verb|\d+?| to expand its match by consuming \verb|8|.
Then, \verb|x| consumes the first \verb|x| in the target string.
After that, the end-of-line matcher (\verb|$|) fails, forcing the \verb|x| to release the character again.
Now, \verb|\d+?| is asked to expand its initial match, which fails, resulting in an overall fail.
It is important to realize that the lazy quantifier always expands its match until the succeeding \Verb|x| can match the first time, but never beyond, no matter to which target string the example \textsc{RegEx} is applied to.
The reason is that the two languages defined by \verb|\d+?| and \verb|x| do not intersect. 
Hence, if the literal matcher releases its matched character, it can never be consumed by the digit matcher.
In other words, the result does not change, when the lazy quantifier is replaced by a possessive one, making it safe to convert to DPL.
In conclusion, the observation from the previous section is not limited to greedy quantifiers, but also applicable to lazy quantifiers.

However, when the matchers intersect, the conversion cannot be safely performed, as shown in Figure~\ref{lst:unsafe-lazy-intersect}.
After \verb|\w+?| consumes the first character \verb|H|, \verb|[a-z]| consumes the second one, resulting in an overall match.
The possessive DPL counterpart would consume all characters of \verb|Hello| until there is no character for the \verb|[a-z]| left to match, resulting in an overall fail.

\paragraph{Last Lazy Quantifier (LLQ).}
If a lazy quantifier appears before an end-of-line matcher (\verb|$|), it always matches until the end of the target string.
Therefore, this conversion is safe as well.
However, if the matcher appears at the very end of the pattern, different rules apply.
Because no matcher comes afterwards, the last matcher never has to expand beyond its initial match.
In other words, it always matches the minimum required number of characters if possible.
That is, lazy quantifiers \Verb|{x,y}?| at the end of a pattern must be converted to \Verb|{x}|.
Consequently, \Verb|+?| translates to \Verb|{1}|, and \Verb|*?| translates to \Verb|{0}|.
The latter case implies that the entire matcher never consumes any characters, so it can be omitted.

\paragraph{Last Successor Lazy Quantifier (SLQ).}
Lastly, one special case applies to the dot-matcher in combination with a lazy quantifier, such as \verb|.+?| and \verb|.*?|.
If such a matcher is followed by exactly one matcher, \ie its successor is the last matcher in the pattern, it can be safely converted to the \textit{lazy-like} counterpart \verb|LD|.
As shown in Figure~\ref{lst:safe-lazy-dot}, the \textsc{RegEx} behaves the same way as its DPL equivalent \Verb|LD+ "!"| would.
Note that any additional matcher may break this behavior.
For example, adding a \Verb|EOL| (respective \verb|$|) matcher at the end, would require the \verb|LD| to expand its match, which is not supported, resulting in an overall fail.
Note that multiple literal matchers are considered one single matcher in the DPL, as mentioned in \secref{sec:dpl-syntax}.
Hence, this strategy is also applicable to patterns where a dot-matcher is followed by multiple literal matchers, such as \Verb|.+?abc|, which translates to \Verb|LD+ "abc"|.

\subsection{Pattern Validation (Step 2)}
\label{sec:validation}

The original \textsc{RegEx} and the generated DPL pattern are considered semantically equivalent only if their defined languages are equivalent. 
That is, the DPL pattern must accept all strings that are accepted by the \textsc{RegEx} and reject all strings that are rejected by the \textsc{RegEx}. 
However, when dealing with \textsc{RegExes}, it is often impossible to enumerate all strings of the defined language, if it is infinite (\eg when unbound quantifiers such as \verb|*| or \verb|+| occur). 
One way to test the equivalence of two regular expressions is to compare their minimal deterministic finite automata (DFA) representations~\cite{Almeida2009}.
However, this cannot be used in our approach, because no algorithm exists that converts DPL to DFA. 

To overcome this, we adopt the approach proposed in~\cite{Zhong2018}, randomly generating a fixed number of strings based on the original \textsc{RegEx}. 
These generated strings then serve as positive test cases for the DPL pattern.
Negative test cases can be obtained by generating matching strings based on the \textsc{RegEx}'s complement. 
Those negative tests pass if they are \textit{not matched} by the converted DPL pattern. 
Although such test cases cannot formally proof semantic equivalence, they provide strong empirical evidence (similar to unit tests). 
In the remainder of the paper, we name a pattern translation from \textsc{RegEx} to DPL that passed these criteria as a "safe" translation.

All test cases are finally matched against the converted DPL pattern.
A positive test case is considered to be passed, only if both \begin{enumerate*}[label=(\roman*)]
    \item it matches the string entirely
    \item it extracts the same field values as the original \textsc{RegEx} 
\end{enumerate*}.

\subsection{Pattern Optimization (Step 3)}
\label{sec:optimization}

After a \textsc{RegEx} was converted to DPL and validated, optimization of the DPL pattern can be initiated. 
That is, potential high-level DPL matchers ($\hookrightarrow$~\secref{sec:high-level-matchers}) are detected.

\subsubsection{Optimization Objective}
\label{sec:opti-objective}

One potential solution to detect DPL's high-level matchers is to map each of them to a corresponding \textsc{RegEx} and search for this sub-pattern in the original \textsc{RegEx}. 
However, it is unlikely that an exact match can be found. 

\figref{lst:rc} presents two illustrative \textsc{RegEx} examples. %
Both examples extract an IP port, which can be matched by \verb|INT| in DPL. 
Line 2 illustrates a variant that precisely matches one to five digits, which is less than what would be matched by the \verb|INT| matcher. 
Line 5 depicts a variant that matches an arbitrary number of digits, which is more than what would be matched by the \verb|INT| matcher. 
Nevertheless, it is desired to use the \verb|INT| matcher in both cases for the reasons mentioned in \secref{sec:dpl}.

\begin{figure}[h]
    \caption{\textsc{RegEx} matching an IP port}
    \label{lst:rc}
    \begin{Verbatim}[numbers=none,frame=single]
1 // matching less than INT
2 port\s(?<ip_port>\d{1,5})
3
4 // matching more than INT
5 port\s(?<ip_port>\d*)
    \end{Verbatim}
    
\end{figure}

Purely algorithmic detection based on the defined language of the high-level matcher therefore does not seem feasible. 
Instead, other factors shall be included in the detection:

\begin{itemize}
    \item The name of a named capturing group. For example, an export name \verb|src_addr| may indicate that the group extracts an IP address. Note that these names are user-defined and may appear in languages other than English.
    \item The context of the \textsc{RegEx}. For example, if a \textsc{RegEx} contains the literal matcher \verb|HTTP/1\.1|, an export name \verb|reponse_code| can most likely be interpreted as an HTTP response code.
\end{itemize}

\begin{figure}[h]
\caption{Language comparison of a \textsc{RegEx} and DPL pattern before (left) and after (right) the optimization step}
\label{fig:venn-opti}
\centering
\includesvg[width=\columnwidth]{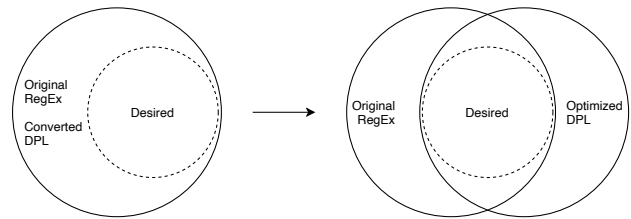}

\end{figure}

We hypothesize that a machine learning (ML) based approach could yield useful results to effectively identify the context of a pattern and the dependencies between the matchers to suggest potential high-level DPL matchers. 
For each prediction, the user must manually review the suggestions and confirm their suitability for the intended use. 
This is necessary because the semantics of the DPL pattern most likely changes when using high-level matchers.
To emphasize this, \figref{fig:venn-opti} compares the language of an original \textsc{RegEx} with the language of its converted DPL pattern before and after detection of high-level matchers.
Even in the event that the languages of both \textsc{RegEx} and DPL pattern are semantically equivalent (see left part of \figref{fig:venn-opti}, \ie a "perfect" translation), high-level matchers can still be employed. 
After this, the defined languages of the \textsc{RegEx} and DPL pattern may differ (right part of \figref{fig:venn-opti}).
However, this is acceptable as long as the optimized DPL pattern continues to include the \textit{desired} language, which represents the \textsc{RegEx} author's true intent.
Note that a \textsc{RegEx} often is an approximation of this \textit{desired} language~\cite{Michael2019}. 

\subsubsection{Matcher Prediction}
\label{sec:prediction}

To detect the high-level DPL matchers as explained above, the basic DPL pattern as a result of the rule-based translation is sent together with the problem and task to \mbox{GPT-4}. 

One advantage of GPT-4 is that the model is already pre-trained~\cite{OpenAI2023} and can understand inputs in natural language as well as programming languages and deliver results without fine-tuning. 
OpenAI does not provide specific information about the training data used. 
Instead, it is vaguely stated that it includes "publicly available data (such as internet data)"~\cite{OpenAI2023}. 
Consequently, the training dataset may also contain the public DPL documentation~\cite{Dynatrace2024b}. 
However, our initial interactions with the model have shown that GPT-4 is not able to generate valid DPL patterns from a target description alone (\eg "Generate a DPL pattern that matches...").
Therefore, we chose to use the \textit{zero-shot prompting} strategy describing the optimization task in detail. 
This technique can lead to more robust results compared to prompts containing demonstrations~\cite{Brown2020}. 

The prompt consists of the following four sections: 
First, the basic DPL pattern is presented as a whole and separated into \textit{fragments}.
We use the term \textit{fragments} to refer to the individual matchers, alternatives, and groups on the highest level of a pattern.
For example, \verb$"a"*"b"("c"|"d")?$ consists of three fragments \verb|"a"*|, \verb|"b"|, and \verb$("c"|"d")?$.
Then, the optimization task is described including a list of high-level matchers which can be suggested for each fragment.
To reduce complexity (also in regards to evaluation), the list of included high-level matchers is limited to the  selection of five specific and generic matchers listed below\footnote{A complete list of built-in matchers is available online~\cite{Dynatrace2024m}.}.
These matchers were selected, because we observed that real-world application logs often contain recurring information such as IP addresses, timestamps, IP ports, HTTP response codes, etc. which are covered by these DPL matchers. 

\begin{itemize}
\setlength\itemsep{0em}
    \item \verb|IPADDR| (matching IPv4 and IPv6 addresses)
    \item \verb|INT|, \verb|LONG|, \verb|DOUBLE|
    \item \verb|TIMESTAMP| (with default format \Verb|yyyy-MM-dd HH:mm:ss|)
\end{itemize}

Next, criteria are defined, based on which matchers can be predicted, namely a matcher's export name and its accepted language.
It is explained that parentheses can be omitted if an entire group is replaced by a matcher and where the export name must be attached.
Lastly, the response format is restricted to JSON and the expected response schema is defined.
Each prompt is preceded with the following message: \textit{"You act as a backend suggesting optimizations for the DPL (Dynatrace Pattern Language) responding in plain JSON."}
The full prompt is provided in a public GitHub repository~\cite{Fragner2024}.

\subsection{\textsc{Reptile} Frontend (Step 4)}

The Frontend component, implemented solely for validation purposes, displays conversion and validation results. 
It also facilitates the review process during pattern optimization.
\begin{figure*}
\caption{\textsc{Reptile}'s frontend showing a converted pattern including high-level matcher suggestions after pattern optimization}
\label{fig:reptile-optimize}
\centering
\includegraphics[width=\textwidth]{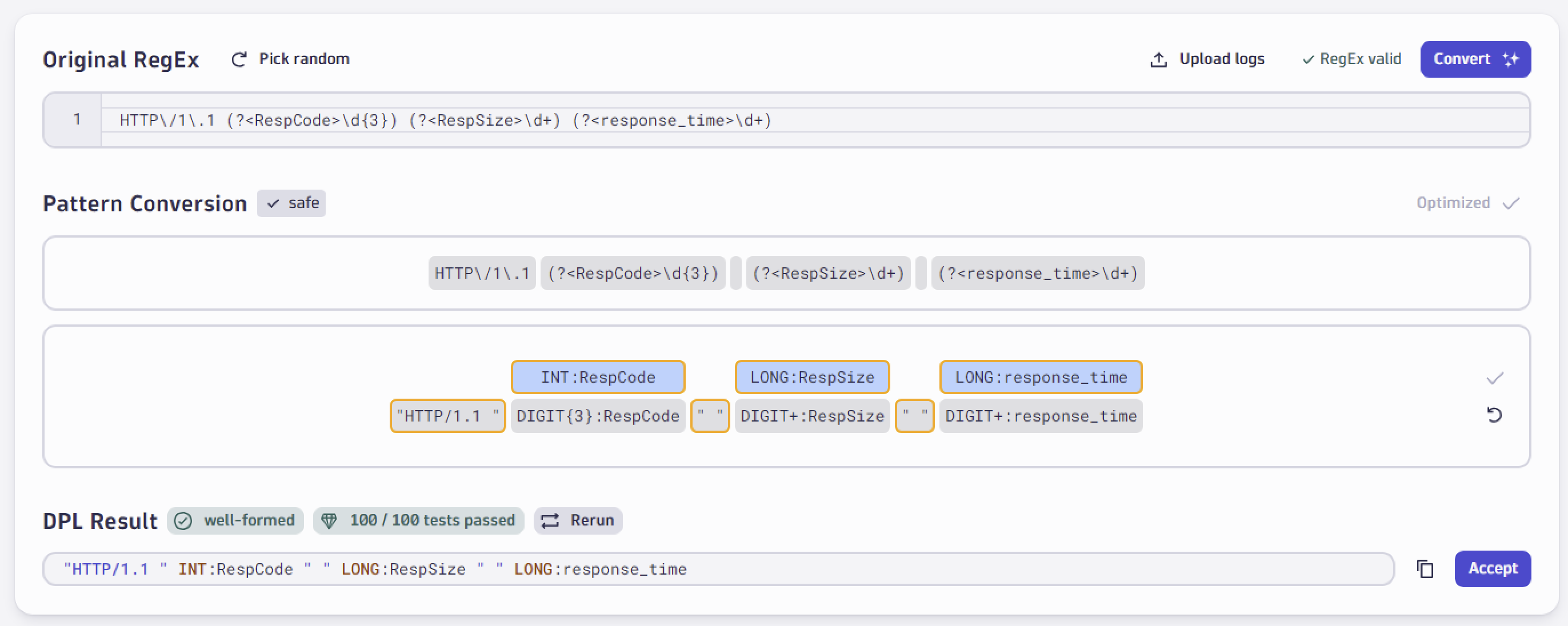}
\end{figure*}
\figref{fig:reptile-optimize} shows a screenshot of \textsc{Reptile}'s frontend\footnote{We also provide a screencast to illustrate the tool and its frontend \url{https://video.dynatrace.com/watch/oxUekmeEehmGHTAZ29d8HN?}}.
The \textsc{RegEx} for conversion can be entered in the topmost input field.
Then, the "Convert" button in the top right initiates the rule-based conversion.
After conversion is finished, the generated DPL pattern is shown in the input field on the bottom where the result can be modified if necessary.
Above, the results of the verification and validation steps are shown as explained in \secref{sec:validation}.
Test results can be analyzed in detail, by clicking the "... tests passed" chip, which opens a sheet overlay containing the test results.

For visibility, the pattern fragments ($\hookrightarrow$~\secref{sec:prediction}) are shown in the central component labeled "Pattern Conversion". 
When the mouse cursor is moved over a fragment, both the fragment and its \textsc{RegEx} or DPL counterpart are highlighted. 
This way, the user can trace how each fragment of the \textsc{RegEx} was converted into which fragment of the DPL pattern.
This is most relevant in case of a best-effort conversion (shown as an "unsafe" conversion in the UI).
That is, the affected greedy or lazy quantifier is transformed into its corresponding possessive DPL quantifier.
In this case, the respective fragment's background color is changed to yellow.
Moreover, the reason why a conversion could not be safely done is shown for the respective fragment, \eg because the quantified matcher intersects with its successor.
Based on this information, the user can decide whether to accept or modify the pattern.

Optionally, the user can initiate the optimization of the DPL pattern.
Each suggested alternative is then rendered above its respective fragment.
The user can decide which alternative to choose by clicking on the fragments.
The final DPL pattern in the bottom line is updated after every selection.
After each update, the syntax is checked automatically.
Generating and executing the test cases can be triggered manually via "Run tests" or "Rerun" buttons respectively.

\subsection{Implementation Details}
\label{sec:impl-details}

We created \textsc{Reptile} as a custom React single-page application (SPA) that uses two services from the Dynatrace environment to validate DPL patterns.
To verify the DPL patterns' syntax, the \textit{Query Assistance Service} is used.
To validate the semantics, the pattern is matched against the test cases described in \secref{sec:validation}  with the help of the \textit{Pattern Developer Service}.
Furthermore, the \textit{App Toolkit}~\cite{Dynatrace2024f} facilitates rapid app development, setup, and deployment, while the \textit{Strato Design System}~\cite{Dynatrace2024g} offers pre-built UI components for creating a contemporary and consistent user interface.
We currently cannot provide a publicly available version of the prototype because it is bound to internal services of Dynatrace. However, in the future we will work on minimizing the dependency on the Dynatrace environment to facilitate the usage of our prototype.

To detect the high-level DPL matchers as explained in \secref{sec:optimization}, a dedicated GPT-4 instance without fine-tuning is used. 
It is accessed via the \textit{Azure OpenAI Service}~\cite{Microsoft2024b}.
To detect if the defined languages of an element and its successor intersect ($\hookrightarrow$~\secref{sec:conversion}), we use the \textit{greenery} library~\cite{greenery2024}.
Because this library is written in Python, it cannot be directly used within \textsc{Reptile}.
To overcome this limitation, we deployed \textit{greenery} as an AWS Lambda serverless function.
This library is also used to obtain a \textsc{RegEx}'s complement necessary to generate negative test cases ($\hookrightarrow$~\secref{sec:validation}).
The \textit{reregexp} library~\cite{reregexp2018} is used to generate random strings based on the original \textsc{RegEx} and its complement, which serve as positive and negative test cases as described in \secref{sec:validation}.

\section{Evaluation}
\label{sec:evaluation}

This section evaluates the \textsc{Reptile} approach.
First, \secref{sec:eval-conversion} assesses the proportion of safely convertible \textsc{RegExes}. %
Please note, as described above, we use the term "safe" conversion if a \textsc{RegEx} can be converted into DPL while retaining the original semantics.
Second, \secref{sec:eval-optimization} evaluates the accuracy of \textsc{Reptile}'s pattern optimization.

\subsection{Rule-Based Conversion}
\label{sec:eval-conversion}

To answer \hyperref[rq:2]{RQ2}, this section assesses the efficacy of our approach. 
To this end, we utilized the real-world dataset from \secref{sec:feature-count} to assess the number of \textsc{RegExes} affected by our conversion strategies.
The program for counting the \textsc{RegEx} features was extended by a new feature category for every conversion strategy presented in \secref{sec:conversion}.

Tables \ref{tbl:greedy-quantifier-results}~and~\ref{tbl:lazy-quantifier-results} show the coverage of the quantifier conversion strategies presented in \secref{sec:conversion}. Overall, 2,640 greedy quantifiers were found in 893 out of 946 \textsc{RegExes} of our industrial dataset. The majority of greedy quantifiers, namely 1,546 (58.56\%), are NGQs occurring in 386 \textsc{RegExes}. Note, while LGQs did not occur most frequently, they affected more \textsc{RegExes}, namely 550. Regarding greedy quantifiers, 658 (\ie 893 - 235) out of 893 (= 73.7\%) affected \textsc{RegExes} were safely converted with \textsc{Reptile}. 235 \textsc{RegExes} contained at least one greedy quantifier for which \textsc{Reptile} does not provide a conversion strategy.

In contrast to greedy quantifiers, only 177 lazy quantifiers were found in 63 out of 946 \textsc{RegExes}. The majority of them, namely 41, are SLQs occurring in 41 \textsc{RegExes}. Our industrial dataset did not contain any FLQs or LLQs. Regarding lazy quantifiers, 42 (\ie 63 - 21) out of 63~(=~66.7\%) affected \textsc{RegExes} were safely converted with \textsc{Reptile}.

\begin{table}
\caption{Greedy quantifier conversion results}
\label{tbl:greedy-quantifier-results}

\resizebox{\columnwidth}{!}{%
\begin{tabular}{ l r r }
\toprule
Type & Total \# & Affected \# \\
\midrule
Fixed Greedy Quantifier (FGQ) & 167 & 94\\
Last Greedy Quantifier (LGQ) & 558 & 550\\
Non-Intersecting Greedy Quantifier (NGQ) & 1546 & 386\\
Remaining & 369 & 235\\
\midrule
Total Greedy Quantifiers & 2640 & 893\\
\bottomrule
\end{tabular}
}
\end{table}

\begin{table}
\caption{Lazy quantifier conversion results}
\label{tbl:lazy-quantifier-results}
\resizebox{\columnwidth}{!}{%
\begin{tabular}{ l r r }
\toprule
Type & Total \# & Affected \# \\
\midrule
Fixed Lazy Quantifier (FLQ) & 0 & 0\\
Non-Intersecting Lazy Quantifier (NLQ) & 8 & 6\\
Last Lazy Quantifier (LLQ) & 0 & 0\\
Last Successor Lazy Quantifier (SLQ) & 41 & 41\\
Remaining & 128 & 21\\
\midrule
Total Lazy Quantifiers & 177 & 63\\
\bottomrule
\end{tabular}
}
\vspace*{-0.2cm}
\end{table}

\begin{table*}
\caption{Conversion results for all conversion strategies applied in combination}
\label{tbl:evaluation-groups}
\begin{tabular}{ l r r r l }
\toprule
Type & Affected \# & Affected \% & Combined \% & Conversion \\
\midrule
Initially converted & 16 & 1.7\% & \multirow{2}{*}{\textbf{73.7\%}} & \multirow{2}{*}{safe} \\
Additionally converted & 681 & \textbf{72.0\%} & & \\
\midrule
Dot-matcher with greedy or lazy quantifier & 163 & 17.2\% & \multirow{2}{*}{26.1\%} & \multirow{2}{*}{best-effort} \\
Remaining greedy quantifiers & 84 & 8.9\% & & \\ 
\midrule
Quantified named capturing groups & 2 & 0.2\% & 0.2\% & not possible \\
\toprule
Total & 946 & 100.0\% & 100.0\% & - \\
\bottomrule
\end{tabular}
\vspace*{1mm}
\vspace*{-0.7cm}
\end{table*}
\tabref{tbl:evaluation-groups} shows the results for all conversion strategies applied in combination.
Column \textit{Affected~\#} corresponds to the number of \textsc{RegExes} covered.
\textit{Affected~\%} reports the same proportion relative to the dataset size (946 \textsc{RegExes}).
The results show that \textsc{Reptile}'s rule-based conversion increased the number of safely converted \textsc{RegExes} by 681 to 697 (= 16 + 681) or 73.7\% safely converted \textsc{RegExes}.

Due to an occurrence of a quantified named capturing group, conversion was syntactically impossible for only 2 \textsc{RegExes}.
For the remaining 26.1\% of the \textsc{RegExes}, \textsc{Reptile} provided a best-effort conversion.
The majority of these (17.2\%) cannot be safely converted, because they contain a quantified dot-matcher (either greedy or lazy).
The remaining 8.9\% contain some other quantified matcher, which in all cases are observed to be greedy.
Note that these \textsc{RegExes} may also contain dot-matchers, but additionally contain at least one other quantified matcher. 

The correctness was validated for the 697 (= 16 + 681) converted \textsc{RegExes}, using randomly generated samples as described in \secref{sec:validation}.
Due to a dependency on the \textit{greenery} library (as discussed in \secref{sec:impl-details}), negative test case generation was only conducted for a subset of \textsc{RegExes}.
The reason for this is that \textit{greenery} is rigorous in its handling of escapings. 
While in PCRE, for example, both \verb|\%| and \verb|%| are recognized as literal matchers, \textit{greenery} only allows the latter variant. 
However, not all special cases are documented in the library. 
To work around this issue and also save resources, we created negative tests for 200 randomly selected \textsc{RegExes} which are accepted by \textit{greenery}.

We performed two test runs with 500 test cases each, resulting in 1,000 samples per \textsc{RegEx}.
All 897,000 generated test cases (697 * 1,000 = 697,000 positive and 200 * 1,000 = 200,000 negative test cases) were reported to be successful.
That is, for positive tests, both the overall matches and their individual captured values were found to be equivalent.
For the negative tests, no match was reported as desired.
In conclusion, these results indicate full semantic congruence and, therefore, correct conversion of \emph{all} converted \textsc{RegExes}. 

\begin{subbox}{Answer to \hyperref[rq:2]{RQ2}}
In a real-world dataset containing 946 \textsc{RegExes}, \textsc{Reptile} increased the number of safely converted \textsc{RegExes} (that is, automatically without human judgment) from initially 1.7\% to 73.7\%.
\end{subbox}

\subsection{Pattern Optimization}
\label{sec:eval-optimization}

This section presents the evaluation of the correctness of \textsc{Reptile}’s DPL pattern optimization to answer the final research question \hyperref[rq:3]{RQ3}. 

\subsubsection{Evaluation Datasets}

For this evaluation, we require a set of \textsc{RegExes} and actual log entries that represent the ground truth to check whether the optimized DPL patterns correctly match the corresponding parts of a log entry. For that, we could not use the \textsc{RegExes} from the previous evaluation because, due to confidentiality reasons, we did not have access to the actual log files from our industrial partner.

To address this issue, we first selected a set of technologies and then searched the internet for corresponding logs and \textsc{RegExes} to parse them. We limited the technologies to logs, such as from Apache, NGINX, AWS, for which \textit{DPL Architect}~\cite{Dynatrace2024d} provides built-in DPL patterns. 
For each technology, we entered "[technology] logs" on Google and perplexity\footnote{\url{https://www.perplexity.ai/}} to search the internet for log files from the respective technology. Next, we entered "[technology] regex"  to search the internet for \textsc{RegExes} with which respective log files can be parsed. 
In total, we collected 23 \textsc{RegExes} from 13 technologies, which are enumerated in \tabref{tbl:eval-dataset}.
The column \textit{Logs~\#} shows the overall number of log entries found for each technology. 
The column \textit{Sources} provides information on the logs' origins. 
The majority of \textsc{RegExes} originates from the \textit{Regex101 community}~\cite{Firas2024b} and the official \textit{ChaosSearch} documentation~\cite{Chaossearch2024}.
Their length ranges from 121 to 787 characters, with an average length of 271.6 characters.
We provide the \textsc{RegExes} and links to the corresponding log files in a public GitHub repository~\cite{Fragner2024}.

For each technology, we first combined all logs into a single file. Next, we executed each \textsc{RegEx} on the corresponding log file to collect all log entries that match the \textsc{RegEx}. These log entries represent the positive test cases. 
Conversely, negative test cases were obtained by filtering the corresponding log file for non-matching log entries.  
The columns \textit{$\varepsilon^+$} and \textit{$\varepsilon^-$} in \tabref{tbl:eval-dataset} show the number of positive and negative test cases respectively. Note, for \textsc{RegExes} with more than 1,000 test cases, the log entries were sampled randomly and limited to 1,000 (marked as~* in \tabref{tbl:eval-dataset}). 

Furthermore note, for seven \textsc{RegExes} (P1a--P1d, P7a, P8a--P8b), we observed that some of the corresponding log entries contain optional elements. 
For example, some log entries contain dashes in places where IP addresses are expected.
These are matched by the original \textsc{RegEx} and its corresponding DPL pattern (\eg via \verb|LD| or \verb|NSPACE|), but not by the high-level matcher \verb|IPADDR|, because a dash is not a valid IP address.
Therefore, we removed these log entries and, consequently, test cases from the datasets leading to the number of positive test cases actually used for the evaluation presented in column $\varepsilon_S^+$ in \tabref{tbl:eval-dataset}. 

\begin{table}
\caption{23 \textsc{RegExes} with their corresponding positive ($\varepsilon^+$), sanitized positive ($\varepsilon_S^+$), and negative ($\varepsilon^-$) test cases; entries marked as~* are randomly sampled and limited to 1,000}
\label{tbl:eval-dataset}
\setlength{\tabcolsep}{3pt}
\resizebox{\columnwidth}{!}{%
\begin{tabular}{ l l l r r r r }
\toprule
\textsc{RegEx} & Technology & Sources & Logs \# & $\varepsilon^+$ & $\varepsilon_S^+$ & $\varepsilon^-$ \\
\midrule
P1a & \multirow{4}{*}{Apache Access} & \multirow{4}{*}{\cite{Chuvakin2010, Elastic2024c}} & \multirow{4}{*}{95,332} & *1,000 & 981 & - \\
P1b & & & & *1,000 & 973 & 470 \\
P1c & & & & *1,000 & 982 & *1,000 \\
P1d & & & & *1,000 & 986 & *1,000 \\
\midrule
P2a & \multirow{2}{*}{Apache Error} & \multirow{2}{*}{\cite{Chuvakin2010}} & \multirow{2}{*}{67,456} & *1,000 & 1,000 & *1,000 \\
P2b & & & & 110 & 110 & *1,000 \\
\midrule
P3a & \multirow{4}{*}{NGINX Access} & \multirow{4}{*}{\cite{Firas2024b}} & \multirow{4}{*}{51,462} & *1,000 & 1,000 & 51 \\
P3b & & & & *1,000 & 1,000 & 51 \\
P3c & & & & *1,000 & 1,000 & - \\
P3d & & & & *1,000 & 1,000 & - \\
\midrule
P4a & OpenSSH & \cite{Firas2024b, Zhu2023} & 2,035 & 525 & 525 & *1,000 \\
\midrule
P5a & \multirow{3}{*}{AWS Cloudfront} & \multirow{3}{*}{\cite{aws2018}} & \multirow{3}{*}{76} & 10 & 10 & 66 \\
P5b & & & & 76 & 76 & - \\
P5c & & & & 76 & 76 & - \\
\midrule
P6a & AWS Route 53 & \cite{aws2024b} & 5 & 5 & 5 & - \\
\midrule
P7a & AWS S3 & \cite{aws2024c} & 10 & 10 & 8 & - \\
\midrule
P8a & \multirow{2}{*}{AWS VPC Flow} & \multirow{2}{*}{\cite{aws2024d,awsNotebook2021,Pluralsight2021}} & \multirow{2}{*}{24} & 24 & 21 & - \\
P8b & & & & 24 & 19 & - \\
\midrule
P9a & Core DNS & \cite{Firas2024b,Datadog2023,Digitalocean2024} & 15 & 15 & 15 & - \\
\midrule
P10a & Log4j & \cite{Firas2024b} & 42 & 42 & 42 & - \\
\midrule
P11a & NGINX Error Log & \cite{Firas2024b} & 27 & 27 & 27 & - \\
\midrule
P12a & SV.NET Service Bus & \cite{Firas2024b} & 22 & 22 & 22 & - \\
\midrule
P13a & HDFS Audit Log & \cite{Firas2024b} & 31 & 31 & 31 & - \\\bottomrule
\end{tabular}
}

\end{table}

\subsubsection{Experimental Set-Up}

Using the dataset from above, we first applied \textsc{Reptile}’s rule-based conversion to convert each of the 23 \textsc{RegExes} to a DPL pattern and optimize it with \textsc{Reptile}’s DPL optimization ($\hookrightarrow$~\secref{sec:optimization}). As already explained in that section, we used an OpenAI's GPT-4 instance via Azure OpenAI Service and our zero-shot prompting strategy for the optimization.

Next, for every optimized DPL pattern we obtained the fragments ($\hookrightarrow$~\secref{sec:prediction}) for which a prediction of a high-level matcher was made. For each predicted high-level matcher in a DPL pattern, we classified it as True Positive (TP) or False Positive (FP) by replacing the fragment by the predicted matcher and execute the pattern on the test cases listed in \tabref{tbl:eval-dataset}. 
If \emph{all} test cases passed, the prediction was categorized as TP, otherwise as FP. 
An exception to this is the \verb|TIMESTAMP| matcher.
Because this matcher is used with its default date format ($\hookrightarrow$~\secref{sec:high-level-matchers}), a prediction was categorized as TP if it was predicted for any fragment matching a timestamp (=~\textit{fragment hit}), even if the default format was incorrect.  

The remaining fragments of a pattern were manually categorized as True Negatives (TN) or False Negatives (FN) as follows. A fragment for which the matcher can be applied but was not predicted was categorized as FN. All other remaining fragments were categorized as TN. These are the fragments for which the matcher could not be applied and was also not predicted by \textsc{Reptile}. 
The manual categorization was performed by the first author and the results were validated by a senior software engineer at Dynatrace, who is responsible for the development
and maintenance of the DPL. Conflicting cases were discussed by both of them and agreement was reached for all cases.

Consider the DPL pattern \verb|LD*:ip ":" WORD+:msg| that consists of the three fragments \verb|LD*:ip|, \verb|":"|, and \verb|WORD+:msg|.
For the sake of argument, assume that \verb|IPADDR| was predicted for the first and last fragment during optimization.
This results in TP=1 (correct prediction of first fragment, assuming all test cases passed) and FP=1 (incorrect prediction of last fragment, assuming at least one test case failed).
Further, FN=0, because there is no additional fragment where \verb|IPADDR| could have been correctly predicted. Lastly, TN=1, because the matcher was not predicted for the remaining middle fragment (which is correct).

Based on the TPs, FPs, FNs, and TNs, the precision, recall, F1-score, and Matthews correlation coefficient (MCC) were calculated for each high-level matcher using the formulas below.
\begin{equation*}
\text{Precision} = \frac{\text{TP}}{\text{TP} + \text{FP}}
\hspace*{1cm}
\text{Recall} = \frac{\text{TP}}{\text{TP} + \text{FN}}
\end{equation*}
\vspace*{0.1cm}

\text{used to obtain the \textit{F1-score} as follows:}

\begin{equation*}
\text{F1-score} = 2 \times \frac{\text{Precision} \times \text{Recall}}{\text{Precision} + \text{Recall}}
\end{equation*}
\vspace*{0.1cm}
\begin{equation*}
\text{MCC} = \frac{\text{TP} \times \text{TN} - \text{FP} \times \text{FN}} {\sqrt{(\text{TP} + \text{FP})(\text{TP} + \text{FN})(\text{TN} + \text{FP})(\text{TN} + \text{FN})}}
\end{equation*}
\vspace*{0.1cm}

\subsubsection{Results}

\tabref{tbl:eval-opt-results-2} shows the results for each high-level matcher across all 23 patterns as described before.
The column $\sum$ shows the total number of fragments in our dataset, namely 579.
Referring to the results, we observe high precision 
values of more than 0.85 across all matchers. 
This is due to the low number of false positives (FPs) ranging between 0 and 5 for the five high-level matchers. 
This means that when a prediction is made by \textsc{Reptile}, it is correct in most cases. 

The recall values indicate that for most fragments for which a high-level matcher is possible, such a matcher is also predicted by \textsc{Reptile}.
We emphasize that a recall of 1.00 was achieved for the three matchers \verb|IPADDR|, \verb|LONG|, and \verb|TIMESTAMP|, indicating that all optimization potential was utilized. 
A relatively low recall of 0.71 is only observed for the \verb|DOUBLE| matcher. 
However, this can be attributed to the limited number of 7 fragments for this matcher, potentially leading to less representative results.

Looking at the F1-score and MCC values in \tabref{tbl:eval-opt-results-2}, we observe values of 0.91 and higher for 4 high-level matchers, whereas the best result (0.98 F1-score and MCC) was obtained for \verb|TIMESTAMP|. Also the F1-score and MCC of 0.83 and 0.84, respectively, for \verb|DOUBLE| indicate good prediction performance of \textsc{Reptile}'s pattern optimization.

\begin{subbox}{Answer to \hyperref[rq:3]{RQ3}}
Optimizing 23 real-world patterns, \textsc{Reptile} shows an average F1-score of 0.91 and an average MCC of 0.92 for high-level matcher prediction across five selected matchers.
\end{subbox}

\begin{table*}
\caption{\textsc{Reptile}'s pattern optimization results for 23 \textsc{RegExes} consisting of 579 pattern fragments
}
\label{tbl:eval-opt-results-2}

\begin{tabularx}{\textwidth}{ X r r r r r r r r r r }
\toprule
Matcher & $\sum$ & TP & FP & FN & TN & Precision & Recall & F1-score & MCC \\
\midrule
\verb|IPADDR| & 579 & 20 & 3 & 0 & 556 & 0.87 & 1.00 & 0.93 & 0.93 \\
\verb|INT| & 579 & 43 & 5 & 3 & 528 & 0.90 & 0.93 & 0.91 & 0.91 \\
\verb|LONG| & 579 & 11 & 2 & 0 & 566 & 0.85 & 1.00 & 0.92 & 0.92 \\
\verb|DOUBLE| & 579 & 5 & 0 & 2 & 572 & 1.00 & 0.71 & 0.83 & 0.84 \\
\verb|TIMESTAMP| (fragment hit) & 579 & 23 & 1 & 0 & 555 & 0.96 & 1.00 & 0.98 & 0.98 \\
\midrule
Average & - & - & - & - & - & 0.91 & 0.93 & 0.91 & 0.92 \\
\bottomrule
\end{tabularx}
\vspace*{1mm}
\end{table*}

\section{Discussion}
\label{sec:discussion}

In this section, we first discuss potential implications of our results on practitioners and researchers. Thereafter, we discuss threats to validity of the results from our experiments.

\subsection{Implications}
The results of our evaluation show that \textsc{Reptile} can automatically and correctly convert 73.7\% of the \textsc{RegExes} to DPL patterns. Furthermore, for the other 26.5\% of \textsc{RegExes}, it provides a best-effort conversion that need to be reviewed by a user. However, for these best-effort conversions, most manual work is eliminated by \textsc{Reptile} because the review is only needed for parts/fragments of the \textsc{RegExes} that could not be converted by one of \textsc{Reptile}'s conversion strategy. 
This clearly shows that \textsc{Reptile} will save users time and effort.

Furthermore, \textsc{Reptile}'s pattern optimization offers potential improvements that only need to be confirmed or rejected by the user. The results from the evaluation show high precision and recall values across five selected high-level matchers. Consequently, if a fragment in a DPL pattern can be replaced by an adequate high-level matcher, \textsc{Reptile} predicts it, and most likely it is the correct one. This helps users to create easier-to-understand and maintain DPL patterns with little extra effort. 

\textsc{Reptile} is currently deployed in a Dynatrace internal environment to test its usefulness. Several internal users started using it --- within 4 months, we recorded 29 user sessions with an average duration of more than 50 minutes. 
The tool was also presented at the \textit{Dynatrace Principal Solution Engineer Summit 2024} in Denver.
We received positive feedback from \textit{Solution Engineers} that \textsc{Reptile} is conceived to be highly useful and is expected to greatly support the conversion process. Both pieces of evidence support the benefits of our approach.

Researchers also benefit from the presented results. First, we provide a detailed description of our approach that allows other researchers to adopt it for converting \textsc{RegExes} to other pattern languages, such as Grok. As mentioned in \secref{sec:impl-details}, we will also work on minimizing the dependency on the Dynatrace environment to allow other researchers to use and extend \textsc{Reptile}. Finally, we provide a publicly available dataset that currently contains 23 real-world \textsc{RegExes} and corresponding log files. It can be used to evaluate future extensions of \textsc{Reptile} or other approaches for converting \textsc{RegExes} to DPL or other pattern languages.

\subsection{Threats to Validity}
We identify the following threats to internal and external validity of our findings. 
Regarding internal validity, we found typos in few \textsc{RegExes} of the industrial dataset. \figref{lst:regex-typo} shows an example where the escaping of the dot in \verb|HTTP\/1.1| is missing. 
This error remained undetected, as the dot-matcher matches \textit{everything}, including the desired literal dot. This resulted in an incorrect count of the matcher as dot-matcher in \secref{sec:feature-count}, despite its intended purpose of being a literal matcher. However, a manual inspection of a random sample of \textsc{RegExes} showed that this issue only affects a few patterns, posing little risk to the validity of the presented results. 

\begin{figure}[h]
    \caption{\textsc{RegEx} containing a dot with missing escaping}
    \label{lst:regex-typo}
    \begin{Verbatim}[numbers=none,frame=single]
HTTP\/1.1 (?<RespCode>[0-9]+) (?<RespSize>[0-9]+)
    \end{Verbatim}
    
\end{figure}

Another threat to internal validity concerns the evaluation of the correct conversion of \textsc{RegExes} to DPL patterns in RQ2. Due to a limitation of \textit{greenery}, negative test cases were not generated for all real-world \textsc{RegExes} ($\hookrightarrow$~\secref{sec:eval-conversion}).
But, we are confident that the results are still valid because positive test cases were generated and passed for all patterns which suggests that the converted patterns fully cover the language of the original \textsc{RegExes}. Furthermore, we performed this evaluation with negative test cases for a random sample of 200 \textsc{RegExes} which are accepted by \textit{greenery}. We view this sample as sufficiently large.

Regarding external validity, we performed the experiments for RQ1 and RQ2 with one industrial dataset from a single organization in a specific industry. Therefore, the results might not generalize to other companies. We mitigated this threat by having our dataset assessed by two experienced solution engineers at Dynatrace.  Both found that the dataset contains \textsc{RegExes} that encompass common use cases for extracting key performance indicators~(KPIs) and other business information, similar to those seen from customers in other business domains, such as banking. Furthermore, the evaluation of \textsc{Reptile}'s pattern optimization demonstrated that our approach indeed generalizes to unseen patterns from various other sources. Regarding RQ3, we mitigated this threat by considering \textsc{RegExes} and log files from 13 different popular technologies. 

\textsc{Reptile} currently supports only the conversion from \textsc{RegExes} to DPL, which limits the generalizability of the results. 
However, as mentioned in \secref{sec:introduction}, DPL is the log parsing pattern language offered by Dynatrace. 
And Dynatrace is a leading company in providing software observability solutions \cite{gartner2024} used by many companies from different business domains. 
Therefore, we can safely assume that DPL is a representative pattern language. In addition, \textsc{Reptile} follows a modular design in which individual components, such as the Converter or the Optimizer ($\hookrightarrow$~\figref{fig:app-overview}), can be extended or replaced. This facilitates the adaption of \textsc{Reptile}'s conversion strategies and optimization approach. We plan such extensions as part of our future work, for instance to convert \textsc{RegExes} to Grok.

Another threat to external validity concerns the evaluation of \textsc{Reptile}'s pattern optimization in RQ3 ($\hookrightarrow$~\secref{sec:eval-optimization}). The evaluation was limited to five common high-level matchers while DPL provides more such matchers. We mitigated this threat by selecting the five matchers based on our observations from the real-world dataset. They showed that IP addresses, timestamps, IP ports, HTTP response codes, and other numbers occur frequently in log entries. In future work, we plan to extend our approach to consider further high-level matchers, such as \verb|HEXINT| or \verb|CREDITCARD|. 

Lastly, \textsc{Reptile}'s pattern optimization was only performed and evaluated with GPT-4 using one prompting strategy which might limit the generalizability of the results. First, we would like to emphasize that the goal of this study was not to find the best-performing LLM and prompting strategy, but to assess the feasibility of LLM-based pattern optimization. Second, we chose GPT-4 because it showed good performance in solving many other software engineering tasks~\cite{toqan2024}. 
Future work will be concerned with exploring other LLMs and deep learning algorithms in addition to 
model fine-tuning~\cite{Microsoft2024d} and other prompt engineering strategies, such as \textit{few-shot prompting}~\cite{Brown2020} or \textit{chain-of-thought prompting}~\cite{Wei2022}.

\section{Related Work}
\label{sec:relwork}

Due to the breadth of its use, the body of research on regular expressions is extensive and in it's entirety beyond the scope of this work. Thus, we focus here on approaches with similar intent --- the generation or conversion from and to regular expressions within a limited set of applications. 

The conversion of regular expressions has been of interest for a longer time starting with Glushkov's automata~\cite{glushkov1961abstract} to convert a \textsc{RegEx} into nondeterministic finite automata (NFA) \cite{ziadi1999optimal, bhargava2011construction, kumar2014novel}. In addition, the conversion to parsing expression grammars (PEG) \cite{ford2004parsing} has been proposed \cite{oikawa2010converting, medeiros2014regexes} as PEGs have a more formal basis compared to the diverging definitions and use of \textsc{RegExes}.

While \textsc{RegExes} are very powerful in many application scenarios such as log parsing~\cite{Zhang2023, He2017}, their creation and maintenance are difficult and cumbersome~\cite{Zhang2023, He2017}. Thus researchers have previously investigated advanced methods for creating \textsc{RegExes}. %
In the context of log parsing, synthesizing \textsc{RegExes} from examples is a very interesting but non-trivial option~\cite{tariq2024automatic}.  Bartoli~et~al.~\cite{Bartoli2014} approached the \textsc{RegEx} synthesis problem by applying \textit{multi-objective genetic programming}~(GP), a paradigm rooted in evolutionary principles~\cite{Koza1994}. Later work \cite{Bartoli2016} provides support for additional features, including alternatives and lookarounds. In a related study, Wang~et~al.~\cite{Wang2016} generated patterns for use in filtering tasks. Very recently, Chen et al. \cite{chen2023data} introduced the concept of semantic regular expressions with a corresponding synthesis approach specifically targeted at data extraction scenarios. Semantic \textsc{RegExes} generalize standard \textsc{RegExes} by considering a type and some additional criterion. For instance, the match needs to be a city (type) in Germany (criterion).

Further work focused on generating more robust \textsc{Reg\allowbreak Exes} \cite{Lee2016, Li2020}. Synthesizing \textsc{RegExes} from natural language has been a promising research direction in recent years \cite{Zhong2018, Park2019, Hahn2022, Wei2022}.  Despite promising results, these approaches are not yet ready for practical use~\cite{Ye2020}. The benchmarks used for validation only contain short \textsc{RegExes} and natural language descriptions with limited vocabulary. Real-world \textsc{RegExes} are longer and more complicated, requiring more complex natural language descriptions. 

A recent study by Zhang~et~al.~\cite{Zhang2023b} demonstrated that high-quality results can be achieved with current large language models (LLMs) even without model fine-tuning. 
The authors criticized the use of \textit{sequence-to-sequence} models in previous works, which generate \textsc{RegEx} patterns character after character from left to right. 
It was recognized that this is not aligned with the order in which \textsc{RegExes} are evaluated during pattern matching, which introduces the risk of syntactically incorrect patterns. 

In contrast, their methodology is based on the concept of \textit{chain-of-thought prompting}~\cite{Wei2022}, which involves constructing multiple prompts in a step-by-step manner. 
This approach has been demonstrated to enhance the performance and interpretability of results produced by pre-trained LLMs. 
In their approach, the authors presented a novel \textsc{RegEx} formulation named \textit{chain-of-inference}, where each chain represents a single sub-\textsc{RegEx}. 
During generation, these sub-problems are solved in the order of pattern matching, which mimics the human way of thinking.
At the same time, no changes to the model or its training process are required. 

To overcome some limitations of the approaches presented above, others have suggested multi-modal approaches \cite{Chen2020, Ye2020}. Here, a synthesis from natural language is the first step with a resulting incomplete pattern. In a second step, provided examples are used to complete the patterns. While promising, in those approaches errors propagate in a way that if the initially generated pattern contains errors, so will the final pattern. Li et al.~\cite{Li2021} compensate for this by first generating complete patterns from natural language and then checking against provided examples and correcting if required. The described two-step approaches bear similarity to our work presented here as also our conversion requires multiple steps to ensure  results with a high conversion rate.

As shown here, the conversion and generation of \textsc{Reg\allowbreak Exes} has been intensively studied but to the best of our knowledge, no approach to the translation between pattern languages --- as we have demonstrated --- exists.

A key issue with \textsc{RegExes} in real-world applications is an effect called \textit{catastrophic backtracking}~\cite{Friedl2006, Fujinami2024, Li2020, Davis2019, Berglund_2014_catastrophicbacktracking} which can occur when matchers with backtracking behavior are used (as it is common in many programming languages). 
This will cause super-linear runtime complexity for the matcher which is a functional issue but could also lead to a vulnerability called Regular Expression Denial of Service (REDoS)~\cite{Crosby2003REDoS}. Besides great care in \textsc{RegEx} design and thorough edge-case testing, approaches to combat this issue have been proposed, often by using static analysis of program code to detect vulnerable \textsc{RegExes}~\cite{Kirrage2013StaticAnalysisREDoS, wustholz2017static, Berglund_2014_catastrophicbacktracking} or by synthesizing safe ones \cite{Li2020}. Languages like \textsc{DPL} that do not strive for universal applicability but are targeted at specific use-cases do not employ backtracking exactly to avoid such issues. In usage scenarios such as the ones sketched in this work, converting \textsc{RegExes} to such a representation has to be done in multiple steps to achieve a high degree of compatibility.

\vspace{-0.4cm}
\section{Conclusion}
\label{sec:conclusion}

In this work, we presented \textsc{Reptile}, an approach that combines a rule-based approach for converting \textsc{RegExes} to DPL patterns with a best-effort approach for cases where a full conversion is not possible. Furthermore, we presented \textsc{Reptile}'s pattern optimization approach to optimize the obtained DPL patterns by predicting high-level matchers. For that, we explored the capabilities of using GPT-4 and a zero-shot prompting strategy. 

The evaluation of \textsc{Reptile}'s rule-based conversion with 946 \textsc{RegExes} collected from a large company showed that \textsc{Reptile} safely converted 73.7\% of them. 
For the remaining 26.3\% \textsc{RegExes} it provided a best-effort conversion that required the input from the user to safely convert them. The evaluation of \textsc{Reptile}'s pattern optimization with 23 other real-world \textsc{RegExes} collected from 13 different, popular technologies showed an average F1-score and MCC above 0.91 across five high-level DPL
matchers. These results have ample practical implications for companies that migrate or migrated to a modern log analytics platform, such as Dynatrace. They can use \textsc{Reptile} to automatically and safely convert their \textsc{RegExes} to DPL patterns. 

In future work, we will extend \textsc{Reptile} to consider other source and target pattern languages, such as Grok. Furthermore, we plan to extend our evaluation to \textsc{RegExes} from other business domains.
Regarding \textsc{Reptile}'s pattern optimization, we plan to extend our approach to consider further high-level matchers, such as \verb|HEXINT| or \verb|CREDITCARD|. Furthermore, we plan to explore other LLMs and prompting strategies, such as \textit{few-shot prompting}~\cite{Brown2020} or \textit{chain-of-thought prompting}~\cite{Wei2022}.

\vspace{-0.4cm}

\section*{Acknowledgements}
We would like to thank Dynatrace Austria GmbH for funding and supporting this project. Furthermore, we would like to thank the industrial partner of Dynatrace to provide the dataset of \textsc{RegExes} used for investigating RQ1 and RQ2. %
We also acknowledge that some diagrams in this work contain icons sourced from \textit{Flaticon}\footnote{\url{https://flaticon.com/}}.

\printcredits

\bibliographystyle{model1-num-names}

\bibliography{references}

\begin{thebibliography}{84}
\expandafter\ifx\csname natexlab\endcsname\relax\def\natexlab#1{#1}\fi
\providecommand{\url}[1]{\texttt{#1}}
\providecommand{\href}[2]{#2}
\providecommand{\path}[1]{#1}
\providecommand{\DOIprefix}{doi:}
\providecommand{\ArXivprefix}{arXiv:}
\providecommand{\URLprefix}{URL: }
\providecommand{\Pubmedprefix}{pmid:}
\providecommand{\doi}[1]{\href{http://dx.doi.org/#1}{\path{#1}}}
\providecommand{\Pubmed}[1]{\href{pmid:#1}{\path{#1}}}
\providecommand{\bibinfo}[2]{#2}
\ifx\xfnm\relax \def\xfnm[#1]{\unskip,\space#1}\fi
\bibitem[{Zhang et~al.(2023)Zhang, Qiu, Castellano, Rifai, Chen, and
  Pianese}]{Zhang2023}
\bibinfo{author}{T.~Zhang}, \bibinfo{author}{H.~Qiu},
  \bibinfo{author}{G.~Castellano}, \bibinfo{author}{M.~Rifai},
  \bibinfo{author}{C.~S. Chen}, \bibinfo{author}{F.~Pianese},
\newblock \bibinfo{title}{{System Log Parsing: A Survey}},
\newblock \bibinfo{journal}{IEEE Transactions on Knowledge and Data
  Engineering} \bibinfo{volume}{35} (\bibinfo{year}{2023})
  \bibinfo{pages}{8596--8614}.
\bibitem[{He et~al.(2017)He, Zhu, Zheng, and Lyu}]{He2017}
\bibinfo{author}{P.~He}, \bibinfo{author}{J.~Zhu}, \bibinfo{author}{Z.~Zheng},
  \bibinfo{author}{M.~R. Lyu},
\newblock \bibinfo{title}{{Drain: An Online Log Parsing Approach with Fixed
  Depth Tree}},
\newblock in: \bibinfo{booktitle}{2017 IEEE International Conference on Web
  Services (ICWS)}, \bibinfo{year}{2017}, pp. \bibinfo{pages}{33--40}.
  \DOIprefix\doi{10.1109/ICWS.2017.13}.
\bibitem[{Fu et~al.(2014)Fu, Zhu, Hu, Lou, Ding, Lin, Zhang, and Xie}]{Fu2014}
\bibinfo{author}{Q.~Fu}, \bibinfo{author}{J.~Zhu}, \bibinfo{author}{W.~Hu},
  \bibinfo{author}{J.-G. Lou}, \bibinfo{author}{R.~Ding},
  \bibinfo{author}{Q.~Lin}, \bibinfo{author}{D.~Zhang},
  \bibinfo{author}{T.~Xie},
\newblock \bibinfo{title}{{Where Do Developers Log? An Empirical Study on
  Logging Practices in Industry}},
\newblock in: \bibinfo{booktitle}{Companion Proceedings of the 36th
  International Conference on Software Engineering}, ICSE Companion 2014,
  \bibinfo{publisher}{Association for Computing Machinery},
  \bibinfo{year}{2014}, p. \bibinfo{pages}{24–33}. \URLprefix
  \url{https://doi.org/10.1145/2591062.2591175}.
  \DOIprefix\doi{10.1145/2591062.2591175}.
\bibitem[{Debnath et~al.(2018)Debnath, Solaimani, Gulzar, Arora, Lumezanu, Xu,
  Zong, Zhang, Jiang, and Khan}]{Debnath2018}
\bibinfo{author}{B.~Debnath}, \bibinfo{author}{M.~Solaimani},
  \bibinfo{author}{M.~A.~G. Gulzar}, \bibinfo{author}{N.~Arora},
  \bibinfo{author}{C.~Lumezanu}, \bibinfo{author}{J.~Xu},
  \bibinfo{author}{B.~Zong}, \bibinfo{author}{H.~Zhang},
  \bibinfo{author}{G.~Jiang}, \bibinfo{author}{L.~Khan},
\newblock \bibinfo{title}{{LogLens: A Real-Time Log Analysis System}},
\newblock in: \bibinfo{booktitle}{2018 IEEE 38th international conference on
  distributed computing systems (ICDCS)}, \bibinfo{organization}{IEEE},
  \bibinfo{year}{2018}, pp. \bibinfo{pages}{1052--1062}.
\bibitem[{Le and Zhang(2023)}]{Le2023}
\bibinfo{author}{V.-H. Le}, \bibinfo{author}{H.~Zhang},
\newblock \bibinfo{title}{{Log Parsing with Prompt-based Few-shot Learning}},
\newblock in: \bibinfo{booktitle}{2023 IEEE/ACM 45th International Conference
  on Software Engineering (ICSE)}, \bibinfo{year}{2023}, pp.
  \bibinfo{pages}{2438--2449}. \DOIprefix\doi{10.1109/ICSE48619.2023.00204}.
\bibitem[{Mi et~al.(2013)Mi, Wang, Zhou, Lyu, and Cai}]{Haibo2018}
\bibinfo{author}{H.~Mi}, \bibinfo{author}{H.~Wang}, \bibinfo{author}{Y.~Zhou},
  \bibinfo{author}{M.~R.-T. Lyu}, \bibinfo{author}{H.~Cai},
\newblock \bibinfo{title}{{Toward Fine-Grained, Unsupervised, Scalable
  Performance Diagnosis for Production Cloud Computing Systems}},
\newblock \bibinfo{journal}{IEEE Transactions on Parallel and Distributed
  Systems} \bibinfo{volume}{24} (\bibinfo{year}{2013})
  \bibinfo{pages}{1245--1255}.
\bibitem[{Vaarandi and Pihelgas(2014)}]{Vaarandi2014}
\bibinfo{author}{R.~Vaarandi}, \bibinfo{author}{M.~Pihelgas},
\newblock \bibinfo{title}{{Using Security Logs for Collecting and Reporting
  Technical Security Metrics}},
\newblock in: \bibinfo{booktitle}{2014 IEEE Military Communications
  Conference}, \bibinfo{year}{2014}, pp. \bibinfo{pages}{294--299}.
  \DOIprefix\doi{10.1109/MILCOM.2014.53}.
\bibitem[{Hamooni et~al.(2016)Hamooni, Debnath, Xu, Zhang, Jiang, and
  Mueen}]{Hamooni2016}
\bibinfo{author}{H.~Hamooni}, \bibinfo{author}{B.~Debnath},
  \bibinfo{author}{J.~Xu}, \bibinfo{author}{H.~Zhang},
  \bibinfo{author}{G.~Jiang}, \bibinfo{author}{A.~Mueen},
\newblock \bibinfo{title}{{LogMine: Fast Pattern Recognition for Log
  Analytics}},
\newblock in: \bibinfo{booktitle}{Proceedings of the 25th ACM international on
  conference on information and knowledge management}, \bibinfo{year}{2016},
  pp. \bibinfo{pages}{1573--1582}.
\bibitem[{{Splunk Inc.}(2023)}]{Splunk2023b}
\bibinfo{author}{{Splunk Inc.}}, \bibinfo{title}{{About Splunk regular
  expressions}},
  \bibinfo{howpublished}{\url{https://docs.splunk.com/Documentation/SCS/current/Search/AboutSplunkregularexpressions}},
  \bibinfo{year}{2023}. \bibinfo{note}{[Online; accessed 2024-02-22]}.
\bibitem[{Zhong et~al.(2018)Zhong, Guo, Yang, Xie, Lou, Liu, and
  Zhang}]{Zhong2018b}
\bibinfo{author}{Z.~Zhong}, \bibinfo{author}{J.~Guo},
  \bibinfo{author}{W.~Yang}, \bibinfo{author}{T.~Xie}, \bibinfo{author}{J.-G.
  Lou}, \bibinfo{author}{T.~Liu}, \bibinfo{author}{D.~Zhang},
\newblock \bibinfo{title}{{Generating Regular Expressions from Natural Language
  Specifications: Are We There Yet?}},
\newblock in: \bibinfo{booktitle}{Workshops at the Thirty-Second AAAI
  Conference on Artificial Intelligence}, \bibinfo{year}{2018}.
\bibitem[{Bartoli et~al.(2012)Bartoli, Davanzo, De~Lorenzo, Mauri, Medvet, and
  Sorio}]{Bartoli2012}
\bibinfo{author}{A.~Bartoli}, \bibinfo{author}{G.~Davanzo},
  \bibinfo{author}{A.~De~Lorenzo}, \bibinfo{author}{M.~Mauri},
  \bibinfo{author}{E.~Medvet}, \bibinfo{author}{E.~Sorio},
\newblock \bibinfo{title}{{Automatic Generation of Regular Expressions from
  Examples with Genetic Programming}},
\newblock in: \bibinfo{booktitle}{Proceedings of the 14th annual conference
  companion on Genetic and evolutionary computation}, \bibinfo{year}{2012}, pp.
  \bibinfo{pages}{1477--1478}.
\bibitem[{Bartoli et~al.(2014)Bartoli, Davanzo, De~Lorenzo, Medvet, and
  Sorio}]{Bartoli2014}
\bibinfo{author}{A.~Bartoli}, \bibinfo{author}{G.~Davanzo},
  \bibinfo{author}{A.~De~Lorenzo}, \bibinfo{author}{E.~Medvet},
  \bibinfo{author}{E.~Sorio},
\newblock \bibinfo{title}{{Automatic Synthesis of Regular Expressions from
  Examples}},
\newblock \bibinfo{journal}{Computer} \bibinfo{volume}{47}
  (\bibinfo{year}{2014}) \bibinfo{pages}{72--80}.
\bibitem[{Bartoli et~al.(2016)Bartoli, De~Lorenzo, Medvet, and
  Tarlao}]{Bartoli2016}
\bibinfo{author}{A.~Bartoli}, \bibinfo{author}{A.~De~Lorenzo},
  \bibinfo{author}{E.~Medvet}, \bibinfo{author}{F.~Tarlao},
\newblock \bibinfo{title}{{Inference of Regular Expressions for Text Extraction
  from Examples}},
\newblock \bibinfo{journal}{IEEE Transactions on Knowledge and Data
  Engineering} \bibinfo{volume}{28} (\bibinfo{year}{2016})
  \bibinfo{pages}{1217--1230}.
\bibitem[{Friedl(2006)}]{Friedl2006}
\bibinfo{author}{J.~E.~F. Friedl}, \bibinfo{title}{{Mastering Regular
  Expressions}}, \bibinfo{edition}{3rd} ed., \bibinfo{publisher}{O'Reilly
  Media, Inc.}, \bibinfo{year}{2006}.
\bibitem[{Li et~al.(2021)Li, Li, Xu, Cao, Chen, Hu, Chen, and Cheung}]{Li2021}
\bibinfo{author}{Y.~Li}, \bibinfo{author}{S.~Li}, \bibinfo{author}{Z.~Xu},
  \bibinfo{author}{J.~Cao}, \bibinfo{author}{Z.~Chen}, \bibinfo{author}{Y.~Hu},
  \bibinfo{author}{H.~Chen}, \bibinfo{author}{S.-C. Cheung},
\newblock \bibinfo{title}{{TransRegex: Multi-modal Regular Expression Synthesis
  by Generate-and-Repair}},
\newblock in: \bibinfo{booktitle}{2021 IEEE/ACM 43rd International Conference
  on Software Engineering (ICSE)}, \bibinfo{organization}{IEEE},
  \bibinfo{year}{2021}, pp. \bibinfo{pages}{1210--1222}.
\bibitem[{Ye et~al.(2020)Ye, Chen, Wang, Dillig, and Durrett}]{Ye2020}
\bibinfo{author}{X.~Ye}, \bibinfo{author}{Q.~Chen}, \bibinfo{author}{X.~Wang},
  \bibinfo{author}{I.~Dillig}, \bibinfo{author}{G.~Durrett},
\newblock \bibinfo{title}{{Sketch-Driven Regular Expression Generation from
  Natural Language and Examples}},
\newblock \bibinfo{journal}{Transactions of the Association for Computational
  Linguistics} \bibinfo{volume}{8} (\bibinfo{year}{2020})
  \bibinfo{pages}{679--694}.
\bibitem[{Chen et~al.(2020)Chen, Wang, Ye, Durrett, and Dillig}]{Chen2020}
\bibinfo{author}{Q.~Chen}, \bibinfo{author}{X.~Wang}, \bibinfo{author}{X.~Ye},
  \bibinfo{author}{G.~Durrett}, \bibinfo{author}{I.~Dillig},
\newblock \bibinfo{title}{{Multi-modal Synthesis of Regular Expressions}},
\newblock in: \bibinfo{booktitle}{Proceedings of the 41st ACM SIGPLAN
  conference on programming language design and implementation},
  \bibinfo{year}{2020}, pp. \bibinfo{pages}{487--502}.
\bibitem[{Li et~al.(2020)Li, Xu, Cao, Chen, Ge, Cheung, and Zhao}]{Li2020}
\bibinfo{author}{Y.~Li}, \bibinfo{author}{Z.~Xu}, \bibinfo{author}{J.~Cao},
  \bibinfo{author}{H.~Chen}, \bibinfo{author}{T.~Ge}, \bibinfo{author}{S.-C.
  Cheung}, \bibinfo{author}{H.~Zhao},
\newblock \bibinfo{title}{{FlashRegex: Deducing Anti-ReDoS Regexes from
  Examples}},
\newblock in: \bibinfo{booktitle}{Proceedings of the 35th IEEE/ACM
  International Conference on Automated Software Engineering},
  \bibinfo{year}{2020}, pp. \bibinfo{pages}{659--671}.
\bibitem[{Davis et~al.(2019)Davis, Michael~IV, Coghlan, Servant, and
  Lee}]{Davis2019}
\bibinfo{author}{J.~C. Davis}, \bibinfo{author}{L.~G. Michael~IV},
  \bibinfo{author}{C.~A. Coghlan}, \bibinfo{author}{F.~Servant},
  \bibinfo{author}{D.~Lee},
\newblock \bibinfo{title}{{Why Aren't Regular Expressions a Lingua Franca? An
  Empirical Study on the Re-use and Portability of Regular Expressions}},
\newblock in: \bibinfo{booktitle}{Proceedings of the 2019 27th ACM Joint
  Meeting on European Software Engineering Conference and Symposium on the
  Foundations of Software Engineering}, \bibinfo{year}{2019}, pp.
  \bibinfo{pages}{443--454}.
\bibitem[{Michael et~al.(2019)Michael, Donohue, Davis, Lee, and
  Servant}]{Michael2019}
\bibinfo{author}{L.~G. Michael}, \bibinfo{author}{J.~Donohue},
  \bibinfo{author}{J.~C. Davis}, \bibinfo{author}{D.~Lee},
  \bibinfo{author}{F.~Servant},
\newblock \bibinfo{title}{{Regexes are Hard: Decision-Making, Difficulties, and
  Risks in Programming Regular Expressions}},
\newblock in: \bibinfo{booktitle}{2019 34th IEEE/ACM International Conference
  on Automated Software Engineering (ASE)}, \bibinfo{year}{2019}, pp.
  \bibinfo{pages}{415--426}. \DOIprefix\doi{10.1109/ASE.2019.00047}.
\bibitem[{Zhang et~al.(2023)Zhang, Gu, Chen, and Shen}]{Zhang2023b}
\bibinfo{author}{S.~Zhang}, \bibinfo{author}{X.~Gu}, \bibinfo{author}{Y.~Chen},
  \bibinfo{author}{B.~Shen},
\newblock \bibinfo{title}{{InfeRE: Step-by-Step Regex Generation via Chain of
  Inference}},
\newblock in: \bibinfo{booktitle}{2023 38th IEEE/ACM International Conference
  on Automated Software Engineering (ASE)}, \bibinfo{year}{2023}, pp.
  \bibinfo{pages}{1505--1515}. \DOIprefix\doi{10.1109/ASE56229.2023.00111}.
\bibitem[{Chapman et~al.(2017)Chapman, Wang, and Stolee}]{Chapman2017}
\bibinfo{author}{C.~Chapman}, \bibinfo{author}{P.~Wang}, \bibinfo{author}{K.~T.
  Stolee},
\newblock \bibinfo{title}{{Exploring Regular Expression Comprehension}},
\newblock in: \bibinfo{booktitle}{2017 32nd IEEE/ACM International Conference
  on Automated Software Engineering (ASE)}, \bibinfo{year}{2017}, pp.
  \bibinfo{pages}{405--416}. \DOIprefix\doi{10.1109/ASE.2017.8115653}.
\bibitem[{{Elasticsearch B.V.}(2024)}]{Elastic2024b}
\bibinfo{author}{{Elasticsearch B.V.}}, \bibinfo{title}{Grok processor},
  \bibinfo{howpublished}{\url{https://www.elastic.co/guide/en/elasticsearch/reference/current/grok-processor.html}},
  \bibinfo{year}{2024}. \bibinfo{note}{[Online; accessed 2024-02-22]}.
\bibitem[{{Datadog Inc.}(2024)}]{Datadog2024b}
\bibinfo{author}{{Datadog Inc.}}, \bibinfo{title}{{Parsing}},
  \bibinfo{howpublished}{\url{https://docs.datadoghq.com/logs/log_configuration/parsing/}},
  \bibinfo{year}{2024}. \bibinfo{note}{[Online; accessed 2024-02-22]}.
\bibitem[{{Dynatrace LLC.}(2024{\natexlab{a}})}]{Dynatrace2024b}
\bibinfo{author}{{Dynatrace LLC.}}, \bibinfo{title}{{Dynatrace Pattern
  Language}},
  \bibinfo{howpublished}{\url{https://docs.dynatrace.com/docs/platform/grail/dynatrace-pattern-language}},
  \bibinfo{year}{2024}{\natexlab{a}}. \bibinfo{note}{[Online; accessed
  2024-02-22]}.
\bibitem[{{Dynatrace LLC.}(2024{\natexlab{b}})}]{Dynatrace2024c}
\bibinfo{author}{{Dynatrace LLC.}}, \bibinfo{title}{Extraction and parsing
  commands},
  \bibinfo{howpublished}{\url{https://docs.dynatrace.com/docs/platform/grail/dynatrace-query-language/commands/extraction-and-parsing-commands}},
  \bibinfo{year}{2024}{\natexlab{b}}. \bibinfo{note}{[Online; accessed
  2024-02-22]}.
\bibitem[{Siegfried et~al.(2024)Siegfried, Byrne, Bangera, and
  Crossley}]{gartner2024}
\bibinfo{author}{G.~Siegfried}, \bibinfo{author}{P.~Byrne},
  \bibinfo{author}{M.~Bangera}, \bibinfo{author}{M.~Crossley},
  \bibinfo{title}{2024 Gartner\textsuperscript{\textregistered} Magic
  Quadrant\texttrademark{} for Observability Platforms},
  \bibinfo{type}{Technical Report}, {Gartner, Inc.}, \bibinfo{year}{2024}.
\bibitem[{Fitzgerald(2012)}]{Fitzgerald2012}
\bibinfo{author}{M.~Fitzgerald}, \bibinfo{title}{{Introducing Regular
  Expressions}}, \bibinfo{publisher}{O'Reilly Media, Inc.},
  \bibinfo{year}{2012}.
\bibitem[{Kleene(1956)}]{Kleene1956}
\bibinfo{author}{S.~C. Kleene}, \bibinfo{title}{{Representation of Events in
  Nerve Nets and Finite Automata}}, \bibinfo{publisher}{Princeton University
  Press}, \bibinfo{year}{1956}, pp. \bibinfo{pages}{3--42}.
  \DOIprefix\doi{doi:10.1515/9781400882618-002}.
\bibitem[{Chomsky(1956)}]{Chomsky1956}
\bibinfo{author}{N.~Chomsky},
\newblock \bibinfo{title}{{Three Models for the Description of Language}},
\newblock \bibinfo{journal}{IRE Transactions on Information Theory}
  \bibinfo{volume}{2} (\bibinfo{year}{1956}) \bibinfo{pages}{113--124}.
\bibitem[{Linz and Rodger(2022)}]{Linz2022}
\bibinfo{author}{P.~Linz}, \bibinfo{author}{S.~H. Rodger}, \bibinfo{title}{{An
  Introduction to Formal Languages and Automata}}, \bibinfo{publisher}{Jones \&
  Bartlett Learning}, \bibinfo{year}{2022}.
\bibitem[{Moseley et~al.(2023)Moseley, Nishio, Perez~Rodriguez, Saarikivi,
  Toub, Veanes, Wan, and Xu}]{Moseley2023}
\bibinfo{author}{D.~Moseley}, \bibinfo{author}{M.~Nishio},
  \bibinfo{author}{J.~Perez~Rodriguez}, \bibinfo{author}{O.~Saarikivi},
  \bibinfo{author}{S.~Toub}, \bibinfo{author}{M.~Veanes},
  \bibinfo{author}{T.~Wan}, \bibinfo{author}{E.~Xu},
\newblock \bibinfo{title}{{Derivative Based Nonbacktracking Real-World Regex
  Matching with Backtracking Semantics}},
\newblock \bibinfo{journal}{Proceedings of the ACM on Programming Languages}
  \bibinfo{volume}{7} (\bibinfo{year}{2023}) \bibinfo{pages}{1026--1049}.
\bibitem[{Hazel(1997)}]{PCRE1997}
\bibinfo{author}{P.~Hazel}, \bibinfo{title}{{PCRE - Perl Compatible Regular
  Expressions}}, \bibinfo{howpublished}{\url{https://pcre.org/}},
  \bibinfo{year}{1997}. \bibinfo{note}{[Online; accessed 2024-07-10]}.
\bibitem[{Fujinami and Hasuo(2024)}]{Fujinami2024}
\bibinfo{author}{H.~Fujinami}, \bibinfo{author}{I.~Hasuo},
\newblock \bibinfo{title}{{Efficient Matching with Memoization for Regexes with
  Look-around and Atomic Grouping}},
\newblock in: \bibinfo{booktitle}{European Symposium on Programming},
  \bibinfo{organization}{Springer}, \bibinfo{year}{2024}, pp.
  \bibinfo{pages}{90--118}.
\bibitem[{Berglund et~al.(2014)Berglund, Drewes, and van~der
  Merwe}]{Berglund_2014_catastrophicbacktracking}
\bibinfo{author}{M.~Berglund}, \bibinfo{author}{F.~Drewes},
  \bibinfo{author}{B.~van~der Merwe},
\newblock \bibinfo{title}{Analyzing catastrophic backtracking behavior in
  practical regular expression matching},
\newblock \bibinfo{journal}{Electronic Proceedings in Theoretical Computer
  Science} \bibinfo{volume}{151} (\bibinfo{year}{2014})
  \bibinfo{pages}{109–123}.
\bibitem[{{Dynatrace LLC.}(2024)}]{Dynatrace2024m}
\bibinfo{author}{{Dynatrace LLC.}}, \bibinfo{title}{Log processing grammar},
  \bibinfo{howpublished}{\url{https://docs.dynatrace.com/docs/platform/grail/dynatrace-pattern-language/log-processing-grammar}},
  \bibinfo{year}{2024}. \bibinfo{note}{[Online; accessed 2024-09-14]}.
\bibitem[{{Splunk Inc.}(2017)}]{Splunk2023c}
\bibinfo{author}{{Splunk Inc.}}, \bibinfo{title}{{About the search language}},
  \bibinfo{howpublished}{\url{https://docs.splunk.com/Documentation/SplunkCloud/latest/Search/Aboutthesearchlanguage}},
  \bibinfo{year}{2017}. \bibinfo{note}{[Online; accessed 2024-03-29]}.
\bibitem[{{Splunk Inc.}(2023)}]{Splunk2023d}
\bibinfo{author}{{Splunk Inc.}}, \bibinfo{title}{rex},
  \bibinfo{howpublished}{\url{https://docs.splunk.com/Documentation/SplunkCloud/latest/SearchReference/Rex}},
  \bibinfo{year}{2023}. \bibinfo{note}{[Online; accessed 2024-03-29]}.
\bibitem[{{The regexpp authors}(2021)}]{regexpp2021}
\bibinfo{author}{{The regexpp authors}}, \bibinfo{title}{regexpp},
  \bibinfo{howpublished}{\url{https://github.com/mysticatea/regexpp}},
  \bibinfo{year}{2021}. \bibinfo{note}{[Online; accessed 2024-03-29]}.
\bibitem[{Su et~al.(2023)Su, Li, Peng, and Chen}]{Su2023}
\bibinfo{author}{W.~Su}, \bibinfo{author}{R.~Li}, \bibinfo{author}{C.~Peng},
  \bibinfo{author}{H.~Chen},
\newblock \bibinfo{title}{{Algorithms for Checking Intersection Non-emptiness
  of Regular Expressions}},
\newblock in: \bibinfo{booktitle}{{Theoretical Aspects of Computing -- ICTAC
  2023}}, \bibinfo{publisher}{Springer Nature Switzerland},
  \bibinfo{year}{2023}, pp. \bibinfo{pages}{216--235}.
\bibitem[{{The greenery authors}(2024)}]{greenery2024}
\bibinfo{author}{{The greenery authors}}, \bibinfo{title}{greenery},
  \bibinfo{howpublished}{\url{https://github.com/qntm/greenery}},
  \bibinfo{year}{2024}. \bibinfo{note}{[Online; accessed 2024-04-19]}.
\bibitem[{Almeida et~al.(2009)Almeida, Moreira, and Reis}]{Almeida2009}
\bibinfo{author}{M.~Almeida}, \bibinfo{author}{N.~Moreira},
  \bibinfo{author}{R.~Reis},
\newblock \bibinfo{title}{{Testing the Equivalence of Regular Languages}},
\newblock in: \bibinfo{booktitle}{Electronic Proceedings in Theoretical
  Computer Science}, volume~\bibinfo{volume}{3}, \bibinfo{year}{2009}, pp.
  \bibinfo{pages}{47--57}. \DOIprefix\doi{10.4204/EPTCS.3.4}.
\bibitem[{Zhong et~al.(2018)Zhong, Guo, Yang, Peng, Xie, Lou, Liu, and
  Zhang}]{Zhong2018}
\bibinfo{author}{Z.~Zhong}, \bibinfo{author}{J.~Guo},
  \bibinfo{author}{W.~Yang}, \bibinfo{author}{J.~Peng},
  \bibinfo{author}{T.~Xie}, \bibinfo{author}{J.-G. Lou},
  \bibinfo{author}{T.~Liu}, \bibinfo{author}{D.~Zhang},
\newblock \bibinfo{title}{{SemRegex: A Semantics-Based Approach for Generating
  Regular Expressions from Natural Language Specifications}},
\newblock in: \bibinfo{booktitle}{Proceedings of the 2018 Conference on
  Empirical Methods in Natural Language Processing},
  \bibinfo{publisher}{Association for Computational Linguistics},
  \bibinfo{year}{2018}, pp. \bibinfo{pages}{1608--1618}. \URLprefix
  \url{https://aclanthology.org/D18-1189}.
  \DOIprefix\doi{10.18653/v1/D18-1189}.
\bibitem[{{OpenAI}(2023)}]{OpenAI2023}
\bibinfo{author}{{OpenAI}},
\newblock \bibinfo{title}{{GPT-4 Technical Report}},
\newblock \bibinfo{journal}{arXiv preprint arXiv:2303.08774}
  (\bibinfo{year}{2023}).
\bibitem[{Brown et~al.(2020)Brown, Mann, Ryder, Subbiah, Kaplan, Dhariwal,
  Neelakantan, Shyam, Sastry, Askell et~al.}]{Brown2020}
\bibinfo{author}{T.~Brown}, \bibinfo{author}{B.~Mann},
  \bibinfo{author}{N.~Ryder}, \bibinfo{author}{M.~Subbiah},
  \bibinfo{author}{J.~D. Kaplan}, \bibinfo{author}{P.~Dhariwal},
  \bibinfo{author}{A.~Neelakantan}, \bibinfo{author}{P.~Shyam},
  \bibinfo{author}{G.~Sastry}, \bibinfo{author}{A.~Askell}, et~al.,
\newblock \bibinfo{title}{{Language Models are Few-Shot Learners}},
\newblock \bibinfo{journal}{Advances in neural information processing systems}
  \bibinfo{volume}{33} (\bibinfo{year}{2020}) \bibinfo{pages}{1877--1901}.
\bibitem[{Fragner(2024)}]{Fragner2024}
\bibinfo{author}{J.~Fragner}, \bibinfo{title}{log-regex},
  \bibinfo{howpublished}{\url{https://github.com/fragjulian/log-regex}},
  \bibinfo{year}{2024}. \bibinfo{note}{[Online; accessed 2024-06-17]}.
\bibitem[{{Dynatrace LLC.}(2024{\natexlab{a}})}]{Dynatrace2024f}
\bibinfo{author}{{Dynatrace LLC.}}, \bibinfo{title}{{Dynatrace App Toolkit}},
  \bibinfo{howpublished}{\url{https://developer.dynatrace.com/reference/app-toolkit/}},
  \bibinfo{year}{2024}{\natexlab{a}}. \bibinfo{note}{[Online; accessed
  2024-04-07]}.
\bibitem[{{Dynatrace LLC.}(2024{\natexlab{b}})}]{Dynatrace2024g}
\bibinfo{author}{{Dynatrace LLC.}}, \bibinfo{title}{Strato design system},
  \bibinfo{howpublished}{\url{https://developer.dynatrace.com/reference/design-system/}},
  \bibinfo{year}{2024}{\natexlab{b}}. \bibinfo{note}{[Online; accessed
  2024-04-07]}.
\bibitem[{{Microsoft}(2024)}]{Microsoft2024b}
\bibinfo{author}{{Microsoft}}, \bibinfo{title}{{Azure OpenAI Service models}},
  \bibinfo{howpublished}{\url{https://learn.microsoft.com/en-us/azure/ai-services/openai/concepts/models}},
  \bibinfo{year}{2024}. \bibinfo{note}{[Online; accessed 2024-05-18]}.
\bibitem[{{The reregexp authors}(2018)}]{reregexp2018}
\bibinfo{author}{{The reregexp authors}}, \bibinfo{title}{reregexp},
  \bibinfo{howpublished}{\url{https://github.com/suchjs/reregexp}},
  \bibinfo{year}{2018}. \bibinfo{note}{[Online; accessed 2024-04-14]}.
\bibitem[{{Dynatrace LLC.}(2024)}]{Dynatrace2024d}
\bibinfo{author}{{Dynatrace LLC.}}, \bibinfo{title}{{DPL Architect}},
  \bibinfo{howpublished}{\url{https://docs.dynatrace.com/docs/platform/grail/dynatrace-pattern-language/dpl-architect}},
  \bibinfo{year}{2024}. \bibinfo{note}{[Online; accessed 2024-03-24]}.
\bibitem[{{The Regex101 community}(2024)}]{Firas2024b}
\bibinfo{author}{{The Regex101 community}}, \bibinfo{title}{{Regex101 Community
  Patterns}}, \bibinfo{howpublished}{\url{https://regex101.com/library}},
  \bibinfo{year}{2024}. \bibinfo{note}{[Online; accessed 2024-05-18]}.
\bibitem[{{ChaosSearch Inc.}(2024)}]{Chaossearch2024}
\bibinfo{author}{{ChaosSearch Inc.}}, \bibinfo{title}{{Regex Support}},
  \bibinfo{howpublished}{\url{https://docs.chaossearch.io/docs/regex-support}},
  \bibinfo{year}{2024}. \bibinfo{note}{[Online; accessed 2024-06-07]}.
\bibitem[{Chuvakin(2010)}]{Chuvakin2010}
\bibinfo{author}{A.~A. Chuvakin}, \bibinfo{title}{{Public Security Log Sharing
  Site}}, \bibinfo{howpublished}{\url{https://www.chuvakin.org/} and
  \url{https://log-sharing.dreamhosters.com/}}, \bibinfo{year}{2010}.
  \bibinfo{note}{[Online; accessed 2024-05-18]}.
\bibitem[{{Elasticsearch B.V.}(2023)}]{Elastic2024c}
\bibinfo{author}{{Elasticsearch B.V.}}, \bibinfo{title}{Elastic examples},
  \bibinfo{howpublished}{\url{https://github.com/elastic/examples/tree/master}},
  \bibinfo{year}{2023}. \bibinfo{note}{[Online; accessed 2024-06-07]}.
\bibitem[{Zhu et~al.(2023)Zhu, He, He, Liu, and Lyu}]{Zhu2023}
\bibinfo{author}{J.~Zhu}, \bibinfo{author}{S.~He}, \bibinfo{author}{P.~He},
  \bibinfo{author}{J.~Liu}, \bibinfo{author}{M.~R. Lyu},
\newblock \bibinfo{title}{{Loghub: A Large Collection of System Log Datasets
  for AI-driven Log Analytics}},
\newblock in: \bibinfo{booktitle}{2023 IEEE 34th International Symposium on
  Software Reliability Engineering (ISSRE)}, \bibinfo{publisher}{IEEE Computer
  Society}, \bibinfo{year}{2023}, pp. \bibinfo{pages}{355--366}. \URLprefix
  \url{https://doi.ieeecomputersociety.org/10.1109/ISSRE59848.2023.00071}.
\bibitem[{Srinivasan(2018)}]{aws2018}
\bibinfo{author}{R.~Srinivasan}, \bibinfo{title}{Sample cloudfront access
  logs},
  \bibinfo{howpublished}{\url{https://github.com/aws-samples/amazon-cloudfront-log-analysis}},
  \bibinfo{year}{2018}. \bibinfo{note}{[Online; accessed 2024-06-07]}.
\bibitem[{{Amazon Web Services Inc.}(2024{\natexlab{a}})}]{aws2024b}
\bibinfo{author}{{Amazon Web Services Inc.}}, \bibinfo{title}{{Public DNS query
  logging}},
  \bibinfo{howpublished}{\url{https://docs.aws.amazon.com/Route53/latest/DeveloperGuide/query-logs.html}},
  \bibinfo{year}{2024}{\natexlab{a}}. \bibinfo{note}{[Online; accessed
  2024-06-07]}.
\bibitem[{{Amazon Web Services Inc.}(2024{\natexlab{b}})}]{aws2024c}
\bibinfo{author}{{Amazon Web Services Inc.}}, \bibinfo{title}{{Amazon S3 server
  access log format}},
  \bibinfo{howpublished}{\url{https://docs.aws.amazon.com/AmazonS3/latest/userguide/LogFormat.html}},
  \bibinfo{year}{2024}{\natexlab{b}}. \bibinfo{note}{[Online; accessed
  2024-06-07]}.
\bibitem[{{Amazon Web Services Inc.}(2024{\natexlab{c}})}]{aws2024d}
\bibinfo{author}{{Amazon Web Services Inc.}}, \bibinfo{title}{Flow log record
  examples},
  \bibinfo{howpublished}{\url{https://docs.aws.amazon.com/vpc/latest/userguide/flow-logs-records-examples.html}},
  \bibinfo{year}{2024}{\natexlab{c}}. \bibinfo{note}{[Online; accessed
  2024-06-07]}.
\bibitem[{{kyhau}(2021)}]{awsNotebook2021}
\bibinfo{author}{{kyhau}}, \bibinfo{title}{{VPC Flow Log examples}},
  \bibinfo{howpublished}{\url{https://kyhau.github.io/aws-notebook/VpcFlowLogs.html}},
  \bibinfo{year}{2021}. \bibinfo{note}{[Online; accessed 2024-06-07]}.
\bibitem[{{Pluralsight LLC.}(2021)}]{Pluralsight2021}
\bibinfo{author}{{Pluralsight LLC.}}, \bibinfo{title}{{Working with AWS VPC
  Flow Logs for Network Monitoring}},
  \bibinfo{howpublished}{\url{https://www.pluralsight.com/cloud-guru/labs/aws/working-with-aws-vpc-flow-logs-for-network-monitoring}},
  \bibinfo{year}{2021}. \bibinfo{note}{[Online; accessed 2024-06-07]}.
\bibitem[{Lentz(2023)}]{Datadog2023}
\bibinfo{author}{D.~Lentz}, \bibinfo{title}{{Key metrics for CoreDNS
  monitoring}},
  \bibinfo{howpublished}{\url{https://www.datadoghq.com/blog/coredns-metrics/}},
  \bibinfo{year}{2023}. \bibinfo{note}{[Online; accessed 2024-06-07]}.
\bibitem[{{DigitalOcean LLC.}(2024)}]{Digitalocean2024}
\bibinfo{author}{{DigitalOcean LLC.}}, \bibinfo{title}{{How to Customize
  CoreDNS for Kubernetes Clusters}},
  \bibinfo{howpublished}{\url{https://docs.digitalocean.com/products/kubernetes/how-to/customize-coredns/}},
  \bibinfo{year}{2024}. \bibinfo{note}{[Online; accessed 2024-06-07]}.
\bibitem[{toq(2024)}]{toqan2024}
\bibinfo{title}{{Coding Assistant}},
  \bibinfo{howpublished}{\url{https://prollm.toqan.ai/leaderboard/coding-assistant}},
  \bibinfo{year}{2024}. \bibinfo{note}{[Online; accessed 2024-10-25]}.
\bibitem[{{Microsoft}(2024)}]{Microsoft2024d}
\bibinfo{author}{{Microsoft}}, \bibinfo{title}{Customize a model with
  fine-tuning},
  \bibinfo{howpublished}{\url{https://learn.microsoft.com/en-us/azure/ai-services/openai/how-to/fine-tuning}},
  \bibinfo{year}{2024}. \bibinfo{note}{[Online; accessed 2024-06-12]}.
\bibitem[{Wei et~al.(2022)Wei, Wang, Schuurmans, Bosma, Xia, Chi, Le, Zhou
  et~al.}]{Wei2022}
\bibinfo{author}{J.~Wei}, \bibinfo{author}{X.~Wang},
  \bibinfo{author}{D.~Schuurmans}, \bibinfo{author}{M.~Bosma},
  \bibinfo{author}{F.~Xia}, \bibinfo{author}{E.~Chi}, \bibinfo{author}{Q.~V.
  Le}, \bibinfo{author}{D.~Zhou}, et~al.,
\newblock \bibinfo{title}{{Chain-of-Thought Prompting Elicits Reasoning in
  Large Language Models}},
\newblock \bibinfo{journal}{Advances in neural information processing systems}
  \bibinfo{volume}{35} (\bibinfo{year}{2022}) \bibinfo{pages}{24824--24837}.
\bibitem[{Glushkov(1961)}]{glushkov1961abstract}
\bibinfo{author}{V.~M. Glushkov},
\newblock \bibinfo{title}{The abstract theory of automata},
\newblock \bibinfo{journal}{Russian Mathematical Surveys} \bibinfo{volume}{16}
  (\bibinfo{year}{1961}) \bibinfo{pages}{1}.
\bibitem[{Ziadi and Champarnaud(1999)}]{ziadi1999optimal}
\bibinfo{author}{D.~Ziadi}, \bibinfo{author}{J.-M. Champarnaud},
\newblock \bibinfo{title}{An optimal parallel algorithm to convert a regular
  expression into its glushkov automaton},
\newblock \bibinfo{journal}{Theoretical computer science} \bibinfo{volume}{215}
  (\bibinfo{year}{1999}) \bibinfo{pages}{69--87}.
\bibitem[{Bhargava and Purohit(2011)}]{bhargava2011construction}
\bibinfo{author}{S.~Bhargava}, \bibinfo{author}{G.~Purohit},
\newblock \bibinfo{title}{Construction of a minimal deterministic finite
  automaton from a regular expression},
\newblock \bibinfo{journal}{International Journal of Computer Applications}
  \bibinfo{volume}{15} (\bibinfo{year}{2011}) \bibinfo{pages}{16--27}.
\bibitem[{Kumar and Verma(2014)}]{kumar2014novel}
\bibinfo{author}{A.~Kumar}, \bibinfo{author}{A.~K. Verma},
\newblock \bibinfo{title}{A novel algorithm for the conversion of parallel
  regular expressions to non-deterministic finite automata},
\newblock \bibinfo{journal}{Applied Mathematics \& Information Sciences}
  \bibinfo{volume}{8} (\bibinfo{year}{2014}) \bibinfo{pages}{95}.
\bibitem[{Ford(2004)}]{ford2004parsing}
\bibinfo{author}{B.~Ford},
\newblock \bibinfo{title}{Parsing expression grammars: a recognition-based
  syntactic foundation},
\newblock in: \bibinfo{booktitle}{Proceedings of the 31st ACM SIGPLAN-SIGACT
  symposium on Principles of programming languages}, \bibinfo{year}{2004}, pp.
  \bibinfo{pages}{111--122}.
\bibitem[{Oikawa et~al.(2010)Oikawa, Ierusalimschy, and
  Moura}]{oikawa2010converting}
\bibinfo{author}{M.~Oikawa}, \bibinfo{author}{R.~Ierusalimschy},
  \bibinfo{author}{A.~Moura},
\newblock \bibinfo{title}{Converting regexes to parsing expression grammars},
\newblock in: \bibinfo{booktitle}{Proceedings of the 14th Brazilian symposium
  on programming languages, SBLP}, volume~\bibinfo{volume}{10},
  \bibinfo{year}{2010}.
\bibitem[{Medeiros et~al.(2014)Medeiros, Mascarenhas, and
  Ierusalimschy}]{medeiros2014regexes}
\bibinfo{author}{S.~Medeiros}, \bibinfo{author}{F.~Mascarenhas},
  \bibinfo{author}{R.~Ierusalimschy},
\newblock \bibinfo{title}{From regexes to parsing expression grammars},
\newblock \bibinfo{journal}{Science of Computer Programming}
  \bibinfo{volume}{93} (\bibinfo{year}{2014}) \bibinfo{pages}{3--18}.
\bibitem[{Tariq and Rana(2024)}]{tariq2024automatic}
\bibinfo{author}{S.~Tariq}, \bibinfo{author}{T.~A. Rana},
\newblock \bibinfo{title}{Automatic regex synthesis methods for english: a
  comparative analysis},
\newblock \bibinfo{journal}{Knowledge and Information Systems}
  (\bibinfo{year}{2024}) \bibinfo{pages}{1--31}.
\bibitem[{Koza(1994)}]{Koza1994}
\bibinfo{author}{J.~R. Koza},
\newblock \bibinfo{title}{{Genetic Programming as a Means for Programming
  Computers by Natural Selection}},
\newblock \bibinfo{journal}{Statistics and computing} \bibinfo{volume}{4}
  (\bibinfo{year}{1994}) \bibinfo{pages}{87--112}.
\bibitem[{Wang et~al.(2016)Wang, Gulwani, and Singh}]{Wang2016}
\bibinfo{author}{X.~Wang}, \bibinfo{author}{S.~Gulwani},
  \bibinfo{author}{R.~Singh},
\newblock \bibinfo{title}{{FIDEX: Filtering Spreadsheet Data using Examples}},
\newblock \bibinfo{journal}{ACM SIGPLAN Notices} \bibinfo{volume}{51}
  (\bibinfo{year}{2016}) \bibinfo{pages}{195--213}.
\bibitem[{Chen et~al.(2023)Chen, Banerjee, Demiralp, Durrett, and
  Dillig}]{chen2023data}
\bibinfo{author}{Q.~Chen}, \bibinfo{author}{A.~Banerjee},
  \bibinfo{author}{{\c{C}}.~Demiralp}, \bibinfo{author}{G.~Durrett},
  \bibinfo{author}{I.~Dillig},
\newblock \bibinfo{title}{Data extraction via semantic regular expression
  synthesis},
\newblock \bibinfo{journal}{Proceedings of the ACM on Programming Languages}
  \bibinfo{volume}{7} (\bibinfo{year}{2023}) \bibinfo{pages}{1848--1877}.
\bibitem[{Lee et~al.(2016)Lee, So, and Oh}]{Lee2016}
\bibinfo{author}{M.~Lee}, \bibinfo{author}{S.~So}, \bibinfo{author}{H.~Oh},
\newblock \bibinfo{title}{{Synthesizing Regular Expressions from Examples for
  Introductory Automata Assignments}},
\newblock in: \bibinfo{booktitle}{Proceedings of the 2016 ACM SIGPLAN
  International Conference on Generative Programming: Concepts and
  Experiences}, \bibinfo{year}{2016}, pp. \bibinfo{pages}{70--80}.
\bibitem[{Park et~al.(2019)Park, Ko, Cognetta, and Han}]{Park2019}
\bibinfo{author}{J.-U. Park}, \bibinfo{author}{S.-K. Ko},
  \bibinfo{author}{M.~Cognetta}, \bibinfo{author}{Y.-S. Han},
\newblock \bibinfo{title}{{SoftRegex: Generating Regex from Natural Language
  Descriptions using Softened Regex Equivalence}},
\newblock in: \bibinfo{booktitle}{Proceedings of the 2019 conference on
  empirical methods in natural language processing and the 9th international
  joint conference on natural language processing (EMNLP-IJCNLP)},
  \bibinfo{year}{2019}, pp. \bibinfo{pages}{6425--6431}.
\bibitem[{Hahn et~al.(2022)Hahn, Schmitt, Tillman, Metzger, Siber, and
  Finkbeiner}]{Hahn2022}
\bibinfo{author}{C.~Hahn}, \bibinfo{author}{F.~Schmitt}, \bibinfo{author}{J.~J.
  Tillman}, \bibinfo{author}{N.~Metzger}, \bibinfo{author}{J.~Siber},
  \bibinfo{author}{B.~Finkbeiner},
\newblock \bibinfo{title}{{Formal Specifications from Natural Language}},
\newblock \bibinfo{journal}{arXiv preprint arXiv:2206.01962}
  (\bibinfo{year}{2022}).
\bibitem[{Crosby(2003)}]{Crosby2003REDoS}
\bibinfo{author}{S.~Crosby},
\newblock \bibinfo{title}{Denial of service through regular expressions},
\newblock \bibinfo{publisher}{USENIX Association},
  \bibinfo{address}{Washington, D.C.}, \bibinfo{year}{2003}.
\bibitem[{Kirrage et~al.(2013)Kirrage, Rathnayake, and
  Thielecke}]{Kirrage2013StaticAnalysisREDoS}
\bibinfo{author}{J.~Kirrage}, \bibinfo{author}{A.~Rathnayake},
  \bibinfo{author}{H.~Thielecke},
\newblock \bibinfo{title}{Static analysis for regular expression
  denial-of-service attacks},
\newblock in: \bibinfo{editor}{J.~Lopez}, \bibinfo{editor}{X.~Huang},
  \bibinfo{editor}{R.~Sandhu} (Eds.), \bibinfo{booktitle}{Network and System
  Security}, \bibinfo{publisher}{Springer Berlin Heidelberg},
  \bibinfo{address}{Berlin, Heidelberg}, \bibinfo{year}{2013}, pp.
  \bibinfo{pages}{135--148}.
\bibitem[{W{\"u}stholz et~al.(2017)W{\"u}stholz, Olivo, Heule, and
  Dillig}]{wustholz2017static}
\bibinfo{author}{V.~W{\"u}stholz}, \bibinfo{author}{O.~Olivo},
  \bibinfo{author}{M.~J. Heule}, \bibinfo{author}{I.~Dillig},
\newblock \bibinfo{title}{Static detection of dos vulnerabilities in programs
  that use regular expressions},
\newblock in: \bibinfo{booktitle}{Tools and Algorithms for the Construction and
  Analysis of Systems: 23rd International Conference, TACAS 2017, Held as Part
  of the European Joint Conferences on Theory and Practice of Software, ETAPS
  2017, Uppsala, Sweden, April 22-29, 2017, Proceedings, Part II 23},
  \bibinfo{organization}{Springer}, \bibinfo{year}{2017}, pp.
  \bibinfo{pages}{3--20}.

\end{thebibliography}

\end{document}